\documentclass[aps,preprint,showpacs,superscriptaddress]{revtex4-1}  
\usepackage{graphicx}
\usepackage{subcaption}
\usepackage{bm}   
\usepackage{color}
\usepackage{amssymb}
\usepackage{amsmath}
\usepackage{float}

\begin{document}
	
	\title[Suppression of extreme events]{Suppression of extreme events and chaos in a velocity-dependent potential system with time-delay feedback}
	
	\author{S. Sudharsan}
	\affiliation{Department of Nonlinear Dynamics, Bharathidasan University, Tiruchirappalli - 620 024, Tamilnadu, India}
	\author{A. Venkatesan}
	\affiliation{PG and Research Department of Physics, Nehru Memorial College (Autonomous), Puthanampatti, Tiruchirappalli - 621 007,  Tamil Nadu, India.}
	\author{P. Muruganandam}
	\affiliation{Department of Physics, Bharathidasan University, Tiruchirappalli - 620 024, Tamilnadu, India}
	\author{M. Senthilvelan}
	\email[Correspondence to: ]{velan@cnld.bdu.ac.in}
	\affiliation{Department of Nonlinear Dynamics, Bharathidasan University, Tiruchirappalli - 620 024, Tamilnadu, India}
	\vspace{10pt}
	
\begin{abstract}
	The foremost aim of this study is to investigate the influence of time-delayed feedback on extreme events in a non-polynomial system with velocity dependent potential. To begin, we investigate the effect of this feedback on extreme events for four different values of the external forcing parameter. Among these four values, in the absence of time-delayed feedback, for two values, the system does not exhibit extreme events and for the other two values, the system exhibits extreme events. On the introduction of time-delayed feedback and varying the feedback strength, we found that extreme events get suppressed as well as get induced. When the feedback is positive, suppression occurs for a larger parameter region whereas in the case of negative feedback it is restricted to the limited parameter region. We confirm our results through Lyapunov exponents, probability density function of peaks, $d_{max}$ plot and two parameter probability plot. Finally, we analyze the changes in the overall dynamics of this system under the influence of time-delayed feedback. We notice that complete suppression of chaos occurs in the considered system for higher values of the time-delayed feedback.
\end{abstract}

	%
	%
	%
	%
	%
	\maketitle
\section{Introduction}
\label{sec:1}
Systems whose time evolution not only depends on the information about the present state of the system but also on the information about their past states are called time-delayed systems. Time-delay is inherently present in many biological, chemical, physical and environmental systems as a result of delay in the processing time of information or due to the delay in the propagation of signal. Further, they appear in amplifiers as a result of fewer swtiching speed {\cite{dvbook}}. Although, these delays are present inherently, they bring about a big changeover in the dynamics of the system when they are introduced externally. In particular, time-delay has been used to stabilize the unstable steady states (USS) and the unstable periodic orbits (UPO) {\cite{Pyragas1992Control,Pyragas1993Control,Balanov2005,Hovel2005}}. 
Several extended methods were proposed {\cite{Socolar1994,Socolar1995,Harrington2004}} for the conversion of the stabilization of the periodic orbits. In addition to this they are also used to control the synchronous power grids {\cite{Botcher2020}}.

As far as the mechanical systems are concerned, time-delay has been found to induce an harmful vibration called regenrative chatter {\cite{Rusinek2014,Rusinek2014Mec}}, Very recently, time-delayed feedback has been used to control the chaotic dynamical behaviour of a periodically stressed nonlinear beams upon an elastic basement {\cite{Kenmogneab2022}}. In addition to this, they also suppress the free vibrations that are produced from a flexible cantilever beam {\cite{delaycontrol}}. Further, it has been recently shown that time-delayed feedback can effectively increase the output power's efficiency in a Duffing type variable energy harvestors {\cite{Jin2021}}. Such are the uses of time-delayed feedback in mechanical system. In our work, we intend to analyse the influence of time-delayed feedback on the extreme events in a well-known non-polynomial mechanical system.

Extreme events are rare events that are found to occur in physical, biological, climatic, electronic and ecological models \cite{Krause2015, Kumarasamy2018, Jentsch2005, Ansmann2013, Reinoso2013, Solli2007, Bodai2011, Dysthe2008, Yukalov2012, Moitra2019, Chaurasia2020}. Their occurrences have been reported in several dynamical systems. To name a few, we cite FitzHugh-Nagumo oscillators, Lin\'eard system, Hindmarsh-Rose model,  micromechanical system, memristor-based Li\'enard system, climatic models, electronic circuits, coupled Ikeda map, network of moving agents, network of Josephson junctions, dispersive wave models, Ginzburg-Landau model and nonlinear Schr\"{o}dinger equation~\cite{Ansmann2013, Karnatak2014, Saha2017, Saha2018, Rings2017, Bialonski2015, Ansmann2016, Mishra2018, Bodai2011, Kingston2017, Kingston2020, Kumarasamy2018, Oliveira2016, Ray2019, Chowdhury2019, Ray2020, Cousins2014, Kim2003,Galuzio2014}. These events also appear in the form of optical rogue waves in lasers and optical fibers \cite{Reinoso2013, Bailung2011, Solli2007, Ganshin2008} and they have also been reported in experiments such as epileptic EEG studies in rodents, annular wave flume, and climatic studies \cite{Pisarchik2018, Toffoli2017, Bodai2011}. 

The problem with the occurrence of extreme events is the cataclysmic consequences that they produce and their disastrous aftermath. Athough the determination of mechanism helps to understand about the origination of these events, it becomes very important to devise control strategies so that the occurrence of these events can be stopped. The developed control measures have to be tested for the suppression of extreme events in dynamical models that mimics the physical systems. Several methods that are used in the chaos theory such as time-delayed feedback {\cite{Suresh2018}}, constant bias, second periodic forcing {\cite{sudharsancons}}, threshold activated coupling {\cite{Ray2019}} have been proposed as viable tools to suppress extreme events that arise in several physical models. Usually the problem with the formulated control strategy is that the constructed mitigation method is highly model dependent and method dependent. This is because one cannot have a unique mitigation method applicable to all the systems or models. So this creates the necessity for having more than one mitigation measures for a given system. 
\par Recently the time-delayed feedback has been used to suppress extreme events in L\'ienard type system {\cite{Suresh2018}}.  On the other hand time-delayed feedback has been found to induce the extreme events in the diode laser system with phase conjugate feedback {\cite{Bosco2013},\cite{Mercier2015}}. In addition to these two, extreme events were also observed in the synchronization manifold in FitzHugh-Nagumo oscillators coupled with two time-delays {\cite{Saha2017}}. This ensures that time-delayed feedback cannot be used as a suppressing tool to all the dynamical models. Motivated by the above, in this work, we investigate the effect of time-delayed feedback in a highly nonlinear system, namely the damped and driven non-polynomial mechanical system. This model ~\cite{Nayfeh1979} is well known for the exhibition of rich nonlinear dynamics \cite{Venkatesan1997,Venkatesan1998}. This model represents the dynamics of a particle in a rotating parabola. In particular, the model can be attributed to a motor bike being ridden in a rotating parabolic well in a circus, centrifugation devices, centrifugal filters and industrial hoppers~\cite{Venkatesan1997, Sanderson1977, Bear1984, Lai1997}. Extreme events in these systems refers to the sudden unbounded erratic motion of the particles. As far as the Centrifuges are concerned, such a sudden abnormal motion of the particles will lead to improper separation of the fluid's components. Hence the centrifuge fails to separate components. Whereas in the case of industrial hoppers and similar other industrial equipments, their main usage is to maintain a constant rate of flow of liquid from one part of the machinery to the other. Extreme events here may refer to the sudden unexpected flow of the particles or fluids deviating from the constant rate. In this case the machinery fails to maintain the constant rate thereby damaging the other parts of the machinery which may even result in explosions. Further, systems with velocity dependent potentials such as mass spectrometers, cyclotrons, magnetrons, railguns, manetoplasmodynamic thrusters and the oscillator version of pion-pion interaction \cite{DELBOURGO1969} are important examples of systems that have velocity dependent potentials. Motivated by all the above facts, in this work, we study the effect of time-delay feedback on the emergence of extreme events and on the overall dynamics of the system. In the industrial hopers, time-delay may be implemented in the form external controllers in the experimental set-up.

Through the investigation, we identify two important impacts of the time-delayed feedback on the considered system. They are $(i)$  the presence of this feedback supresses extreme events present in the system to a great extent and $(ii)$ they suppress the chaotic nature of the system. Although several works have been done on the suppression of chaos \cite{Rajasekar2003}, it is for the first time for a nonpolynomial mechanical system we observe the suppression of both extreme events and chaos due to influence of time-delayed feedback. As far as the suppression of extreme events are concerned majorly extreme events are suppressed due to positive time-delayed feedback while in the case of negative time-delayed feedback, we notice that the suppression of extreme events is restricted to a few parameter regions. Further, for higher values of the positive time-delay feedback, chaotic nature of the system suppresses completely for a very small value of the feedback strength while for negative values of the feedback strength chaos suppresses for a comparitively larger value of the feedback strength.

The paper is organised as follows. In Sec. \ref{sec:2}, we discuss about the model and how time-delayed feeback is introduced. In Sec. \ref{sec:3}, we define the extreme event from dynamical systems point of view and the important properties exhibited by it. In Sec. \ref{sec:4} and Sec. \ref{sec:5}, we discuss about the influence of negative and positive time-delayed feedback respectively. In Sec. \ref{sec:6a}, we study the impact of time-delayed feedback on the overall dynamics of the system. In Sec. \ref{sec:6}, the results are consolidated using a two parameter phase diagram and  the way in which extreme events are influenced by the time-delayed feedback are discussed. Finally, In Sec. \ref{sec:7}, we conclude our work. 

\section{The Model}
\label{sec:2}
We consider a non-polynomial system which describes the motion of a freely sliding particle of unit mass on a parabolic wire ($z = \sqrt{\lambda} x^2$) rotating with a constant angular velocity $\Omega$ ($\Omega^2 = \Omega_0^2 = -\omega_0^2 + g\sqrt{\lambda}$), where $\lambda > 0$ and $\omega_0>0$~\cite{Nayfeh1979}. Here $g$ is the acceleration due to gravity, $1/\sqrt{\lambda}$ is the semi-latus rectum of the rotating parabola, and $\omega_0$ is the initial angular velocity~\cite{Nayfeh1979}. A Lagrangian, $L = \frac{1}{2} \left[ (1 + \lambda x^2) \dot{x}^2 - \omega_0^2x^2 \right]$, can be associated with this velocity dependent potential system \cite{Nayfeh1979}. Here the overdot represents time derivative. With additional damping, external drive and time-delayed feedback, the equation of motion turns out to be
\begin{equation}
	(1+\lambda x^2) \ddot{x} +\lambda x \dot{x}^2 + \omega_0^2 x + \alpha \dot{x} = f \cos \omega t \pm \epsilon x_\tau.
	\label{delay}
\end{equation} 
Here $f$ and $\omega$ are the strength and frequency, respectively, of the external force while $\epsilon$ is the strength of the time-delay feedback and $x_\tau=x(t-\tau)$, where $\tau$ is the time-delay. We consider both positive and negative time-delayed feedback case.

\par In the entire work, the parameters in Eq. (\ref{delay}) are fixed as $\lambda=0.5$, $\omega_0^2 = 0.25$, $\alpha = 0.2$, $\tau=0.1$ and $\omega = 1.0$. The other two parameters, namely  $\epsilon$ and $f$, are varied in the respective analysis. The initial conditions are fixed as $x(0) = 0.1$ and $\dot x (0) = 0.1$ unless otherwise explicitly mentioned. 

Following the methodology given in \cite{trunc}, we expand the time-delay term $x(t-\tau)$ using Taylor series, truncate it to the second order and substitute it back into Eq.~(\ref{delay}). As a result, we obtain the following two coupled first order ordinary differential equations, namely 
\begin{eqnarray}
	\dot{x} &=& y, \nonumber \\
	\dot{y} &=& \frac{f~\mathrm{cos}~\omega t - (\alpha \pm \epsilon\tau)y-\lambda xy^2 -(\omega_0^2\mp\epsilon)x}{1+\lambda x^2 \mp \frac{\epsilon \tau^2}{2}} .
	\label{trunc}
\end{eqnarray}

The methodology mentioned in \cite{trunc} has been especially reported for the calculation of Lyapunov exponents in time-delayed systems. We have verified and confirmed that the dynamics of the systems (\ref{delay}) and (\ref{trunc})	are exactly similar for lower values of $\epsilon$. In our study, since both the suppression of chaos and extreme events occur for lower values of $\epsilon$, the dynamics matches perfectly. The plots presented in Sec.~\ref{sec:6a} are computed using Eq.~(\ref{trunc}). Further, the Lyapunov exponent plots and the two parameter diagrams given in the sections~\ref{sec:4}~\&~\ref{sec:5} have been plotted using the truncated Eq.~(\ref{trunc}). While the remaining plots including the probability plots are generated using the delay differential Eq.~(\ref{delay}). 

The complete bifurcation analysis of the system ($\ref{delay}$) without time-delay feedback ($\epsilon=0.0$, $\tau=0.0$) and the route by which it reaches chaos have already been discussed in Ref.~\cite{Venkatesan1997}. Further, the emergence of extreme events in the system (\ref{delay}) without time-delay has also been studied very recently \cite{Sudharsan2021}. In the present work, we investigate the effect of time-delay feedback on the system and in particular, we answer the prime question, whether the time-delayed feedback suppresses the occurrence of extreme events or not.

\begin{figure}[!ht]
	\begin{center}
		\includegraphics[width=0.7\textwidth]{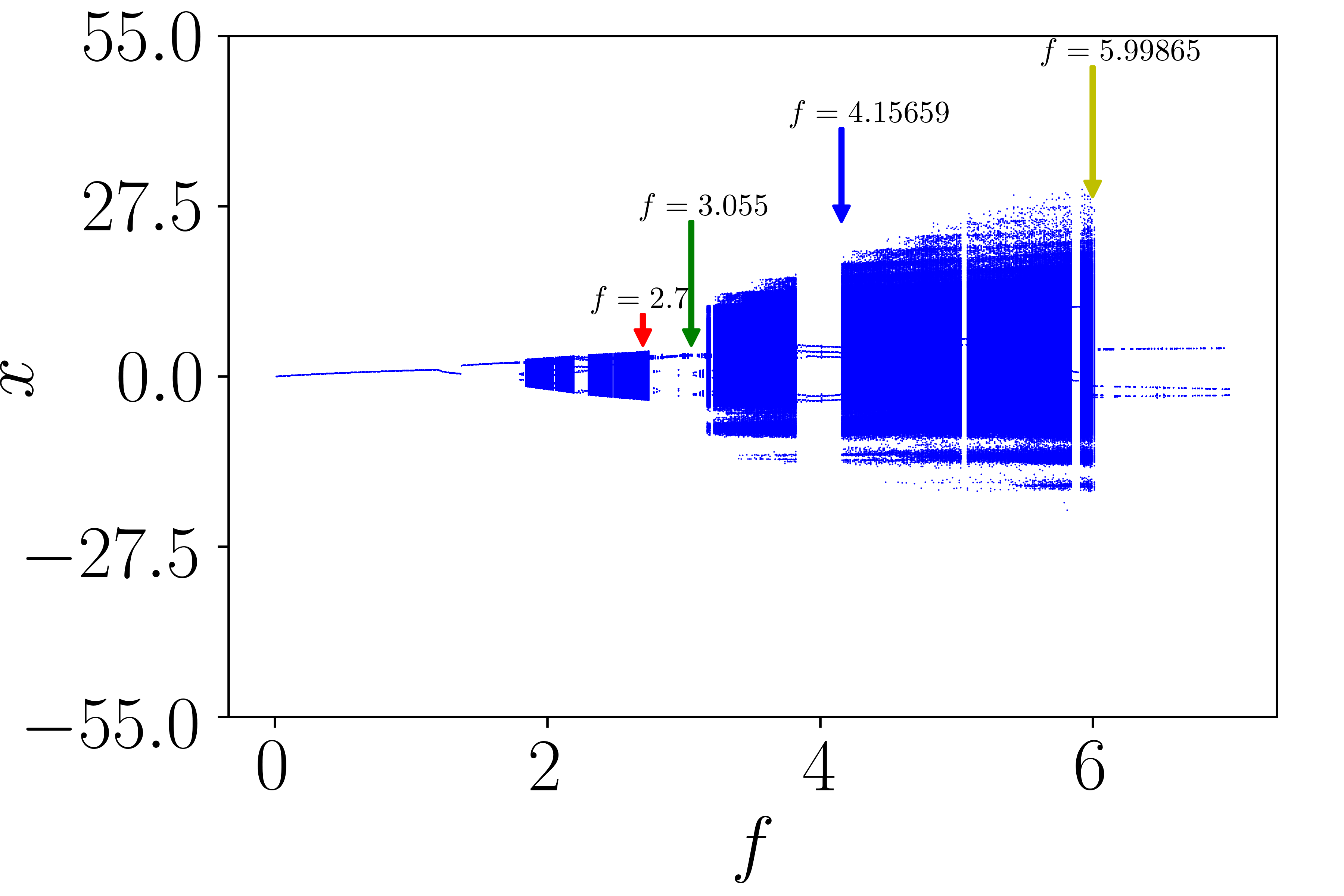}
	\end{center}
	\caption{Bifurcation plot of Eq. (\ref{delay}) with $\epsilon=0.0$ and $\tau=0.0$. The other parameter values are $\lambda=0.5$, $\omega_0^2 = 0.25$, $\alpha = 0.2$ and $\omega = 1.0$. A similar bifurcation diagram using the points at every drive cycles of the external force, for $\epsilon$ up to $7.0$, was reported in \cite{Venkatesan1998}. We use it here to for our analysis.}
	\label{figbif}
\end{figure}

In Fig.~\ref{figbif}, we show the one parameter bifurcation diagram generated by numerically solving Eq.~(\ref{delay}) without time-delay, for different values of $f$. The points in Fig.~\ref{figbif} represent the peak values of $x$ at which $\dot x = 0$. The arrows at $f=2.7,~3.055,~4.15659$ and $5.99865$ are values of the parameter $f$ at which we study the effect of time-delayed feedback on extreme events. It is important to note that, when $\epsilon=0$ and $\tau=0$, the first two values mentioned for $f$ correspond to the point where extreme events do not occur and the last two values correspond to the points where extreme events occur. 

\section{Extreme Events}
\label{sec:3}
In general, extreme events are classified by the presence long tail in distribution of peaks. Although these events are rare in their occurrence, this does not mean that all the rare events correspond to extreme events. Only events that is higher than a certain significant height qualifies to become a rare event. The significant height is also called as qualifier threshold. The threshold of a set of consecutive peaks, say for instance $\{x_i\}$, $i = 1,2,\ldots$, is calculated as $x_{ee}=\langle x \rangle + n\, \sigma_x$, where $\langle x \rangle$ is the mean peak amplitude, $\sigma_x$ is the standard deviation over mean peak amplitude and $n \ge 4$ \cite{Dysthe2008,Jentsch2005,Ray2019}. This threshold value is called as qualifier threshold. The form of qualifier threshold $x_{ee}=\langle x\rangle + n\sigma_x$ has been initially used in the Oceanographic studies to identify rogue waves {\cite{Dysthe2008}} which is an extreme event. Nowadays this formula has been used as an qualifier measure for extreme events in other fields such as optics {\cite{Ray2019}}, mechanical and hydrodynamical systems {\cite{Kumarasamy2018}} and so on. The choice of $n$ varies, depending on the considered system. For instance, for extreme events in Li\'enard system {\cite{Kingston2017}}, $n$ was taken to be equal to $8$, while $n=6$ was fixed for Hindmarsh-Rose model {\cite{Mishra2018}}. For patches of ecological populations, $n=10$ was found to be a good choice {\cite{Chaurasia2020}}.  In this connection, we have used the same qualifier threshold formula in our study.  The value of $n$ that we have fixed is 4, because extreme events in the considered non-polynomial mechanical system have been observed for this choice of $n$ only. From the dynamical systems point of view, crossing of the system's trajectory across this threshold happens whenever a sudden excursion occurs in the trajectory from an already bounded chaotic attractor. The crossing of the system's trajectory over this qualifier threshold $x_{ee}$ can happen even at a very large time. Throughout this work we fix $n=4$ and calculate the value of $x_{ee}$ for very long time iterations, typically of the order of $20\,000\,000$ time units with a step size of $0.01$. This is because, the system which we consider displays a long transient behaviour, some times even until $600\,000$ time units. So in order to avoid the error in the computation of $x_{ee}$, we calculate $x_{ee}$ only after $1\,000\,000$ time units untill $20\,000\,000$ time units. We have also made sure that extreme events occur even for very long runs.   

Further, one of the important characterisations of extreme events is the probability distribution function (PDF) of peaks ($P_n$). This gives us the value of all the peaks and its corresponding probability. The distribution should be fat-tailed such that there is atleast a finite probability of peaks beyond the threshold value. So far in the literature {\cite{Kingston2017}}, in the isolated systems, extreme events have been produced either during the sudden formation of chaotic attractor or due to sudden expansion of a chaotic attractor. So this ensures that all extreme events appearing in single systems are a part of the chaotic attractor. hence while plotting the PDF of peaks for the chaotic non-extreme region, we would normally obtain a Gaussian like distribution in which there would be no peaks beyond the threshold mark (see Fig.~ {\ref{pdf1}(c)}). In a non-extreme chaotic attractor, the trajectories will be bounded due to which the distribution of peaks will take a Gaussian form. Whereas for the chaotic regions with extreme events, we get a fat-tailed or long-tailed distribution and there will be atleast a few peaks beyond the threshold mark (see Fig.~ {\ref{pdf1}}(a) and (b)).  This is because, during the occurrence of extreme events, the trajectories take a sudden long excursion from the otherwise bounded chaotic trajectory. This sudden excursion reflects in the form of long tail in the peak probability distribution function plot. Further, we also compute the value of $d_{max}$ in order to verify the occurrence of extreme events. In the following section, we present our results with necessary characterisation.

\section{Influence of negative time-delayed feedback }
\label{sec:4}
In this section, we study the influence of negative time-delayed feedback on the extreme events through bifurcation plots. Next, we analyse the emergence/suppression of extreme events using appropriate statistical characterizations. The different points at which the extreme events occur and the mechanism by which extreme events occur in the system (\ref{delay}) with $\epsilon=0$ and $\tau=0$ have already been studied in one of our earlier works \cite{Sudharsan2021}. In this work, we study the influence of time-delayed feedback. For this, we choose four different values of $f$, namely $f=2.7$, $f=3.055$, $f=4.15659$ and $f=5.99865$. These points correspond to {\it bounded chaos, period doubling, and two large sized chaotic attractors}. The four points are represented by coloured arrows in Fig. \ref{figbif}. Of these four points, extreme events are found to occur even in the absence of time-delayed feedback only at $f=4.15659$ and $f=5.99865$. We have chosen these four values  of $f$ for which there is a significant effect because of the introduction of the time-delayed feedback. For other values, there was neither emergence nor suppression of extreme events. Now, on the introduction of time-delayed feedback ($\tau=0.1$) and varying the feedback strength $\epsilon$, we observe significant changes in the dynamics of the system at these four values of $f$. To analyse these changes we draw the bifurcation plot with respect to the time-delayed feedback strength $\epsilon$ for each case separately. 

\begin{figure}[!ht]
	\begin{center}
		\includegraphics[width=0.7\linewidth]{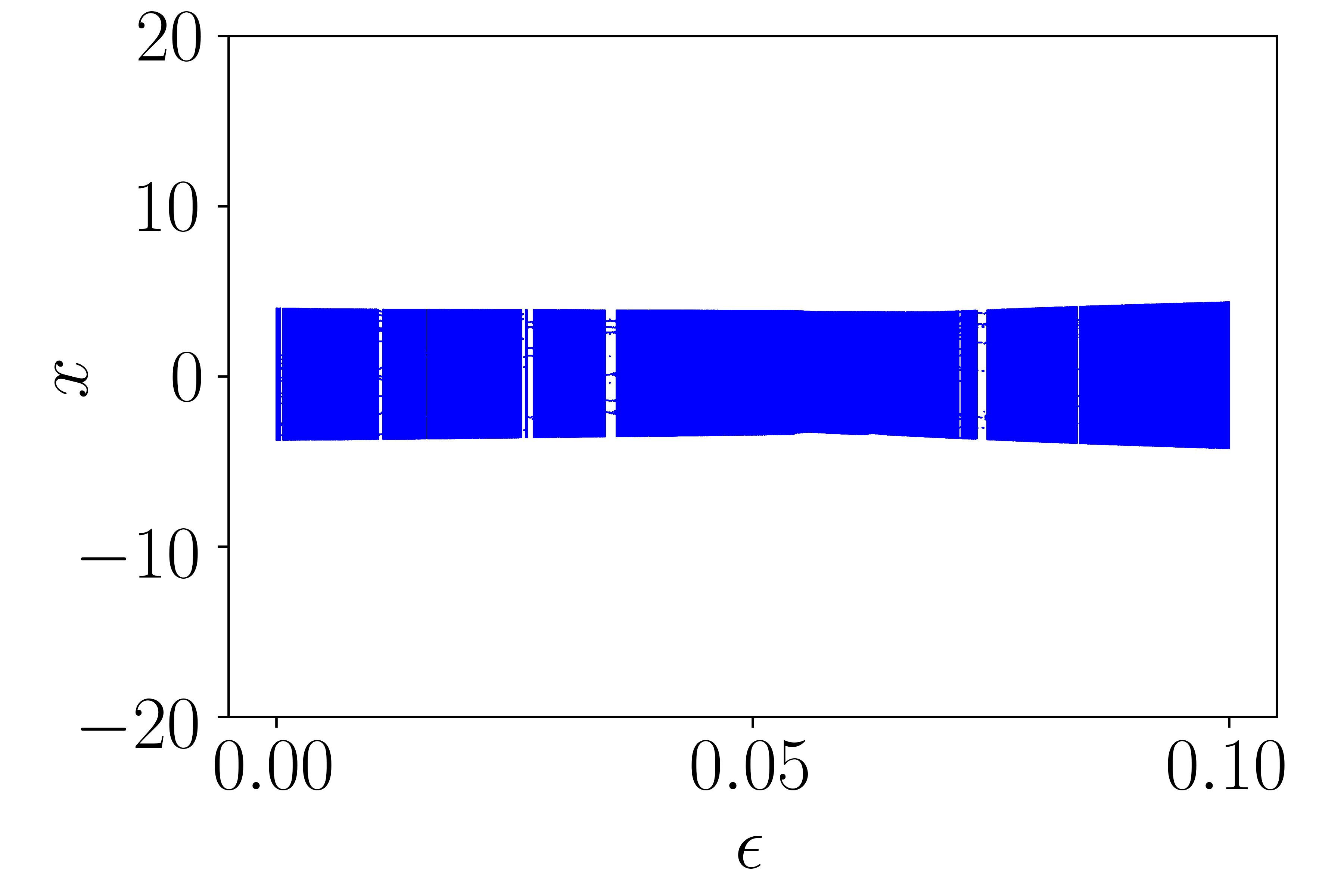}
	\end{center}
	\caption{Bifurcation plot of Eq.~\ref{delay} for $f=2.7$ for varying $\epsilon$}
	\label{bif2.7}
\end{figure}  

At $f=2.7$, the dynamics continues to be chaotic and the size of the chaotic attractor remains almost constant. Only at a few places, tangent bifurcation, period doubling and reverse period doubling occurs. But these periodic dynamics occur only for a very small range of $\epsilon$. After a brief periodic window the dynamics changes back to chaotic nature again. Since no signature of extreme events were found to occur in this parameter space we do not go much into the analysis for this particular $f$. The bifurcation diagram has been displayed just in order to make clear the fact that there are points of $f$ in system (\ref{delay}) where there are no occurrence of extreme events both before and after the introduction of time-delayed feedback eventhough the dynamics is still chaotic.

\begin{figure}[!ht]
	\begin{center}
		\includegraphics[width=0.7\linewidth]{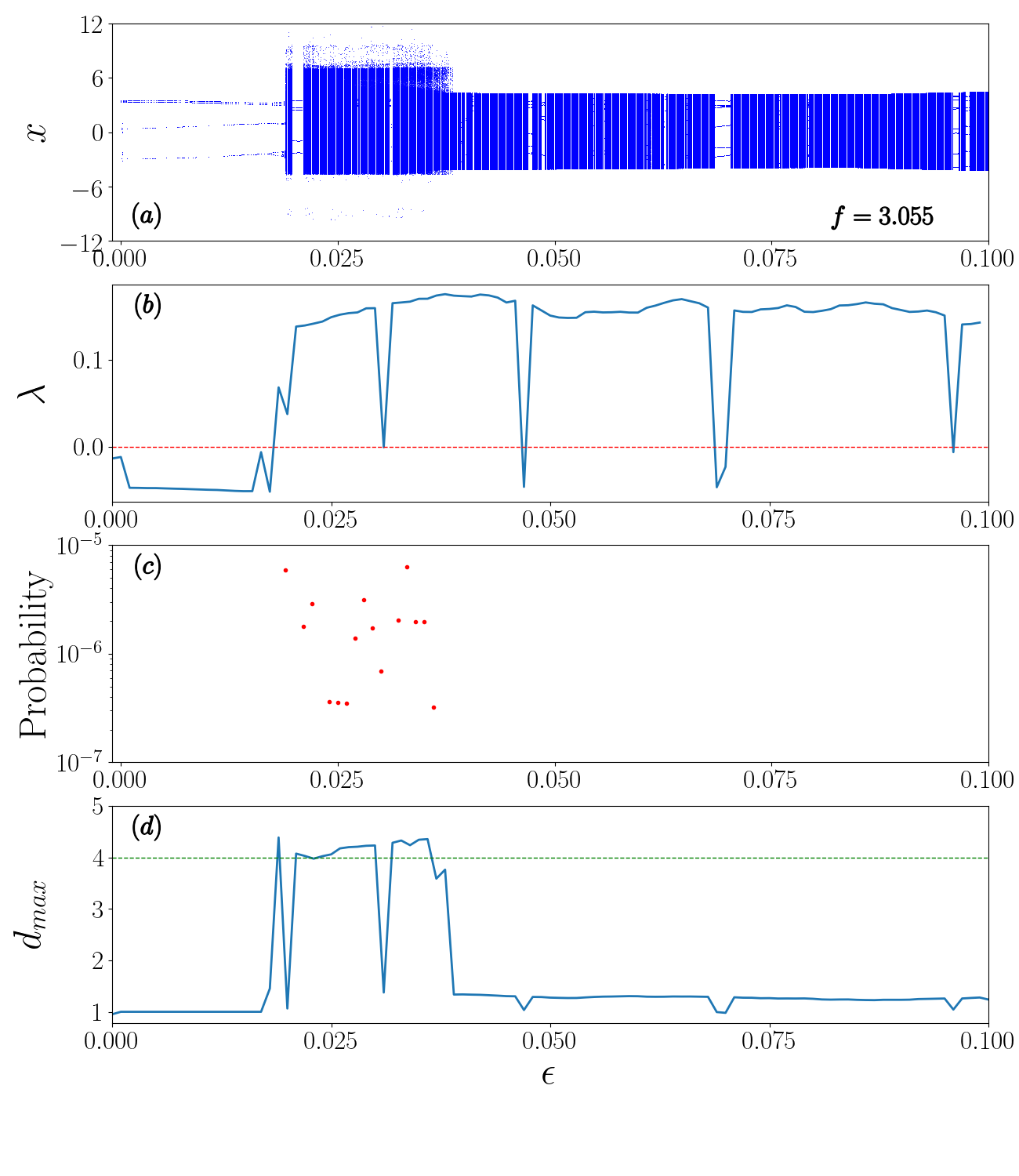}
	\end{center}
	\caption{(a) Bifurcation plot, (b) Largest Lyapunov exponent, (c) Probability plot and (d) $d_{max}$ plot of Eq.~\ref{delay} for $f=3.055$ for varying $\epsilon$}
	\label{bif3.055}
\end{figure}
Now, we check whether the extreme events emerge or not at $f=3.055$ due to the introduction of time-delayed feedback. Here we observe that the dynamics differs much more significantly than the previous case. Before the introduction of time-delayed feedback, system (\ref{delay}) exhibits a multiperiodic dynamics and after the introduction of time-delay and varying the time-delay feedback strength, the system undergoes the following two prominent bifurcations, namely {\it period-doubling} and {\it interior crisis}. Intermittently, tangent bifurcation also occurs at several places giving rise to periodic windows. The nature of periodicity and chaos are verified using the correponding largest Lyapunov exponent. As further, in Fig. \ref{bif3.055}(c), we plot the probability plot for varying $\epsilon$. It can be seen that while increasing the feedback strength, extreme events are induced at $\epsilon=0.019$. This is the point where the system exhibits a sudden expansion of the chaotic attractor. Along with the destruction of chaotic attractor by periodic windows, emergence of extreme events also vary in probability finally getting suppressed completely at $\epsilon=0.037$. This is the point where the reverse interior crisis occurs that is where an expanded chaotic atrractor becomes a bounded chaotic attractor. The fact that chaotic nature prevails in the system even after the suppression of extreme events can be confirmed by the presence of positive Lyapunov exponent beyond $\epsilon=0.037$. But when we further increase the value of $\epsilon$, the chaotic attractor suppresses as dicussed in Sec.~\ref{sec:6a}. The emergence and suppression of the extreme events is further confirmed by the $d_{max}$ plot shown in Fig.~\ref{bif3.055}(d). It can be seen that, while extreme events emerge only when the value of $d_{max}$ is above four and for the regions where no extreme events occur, the value of $d_{max}$ lies below four. 

\begin{figure}
	\centering
	\begin{subfigure}[b]{0.32\textwidth}
		\centering
		\includegraphics[width=\textwidth]{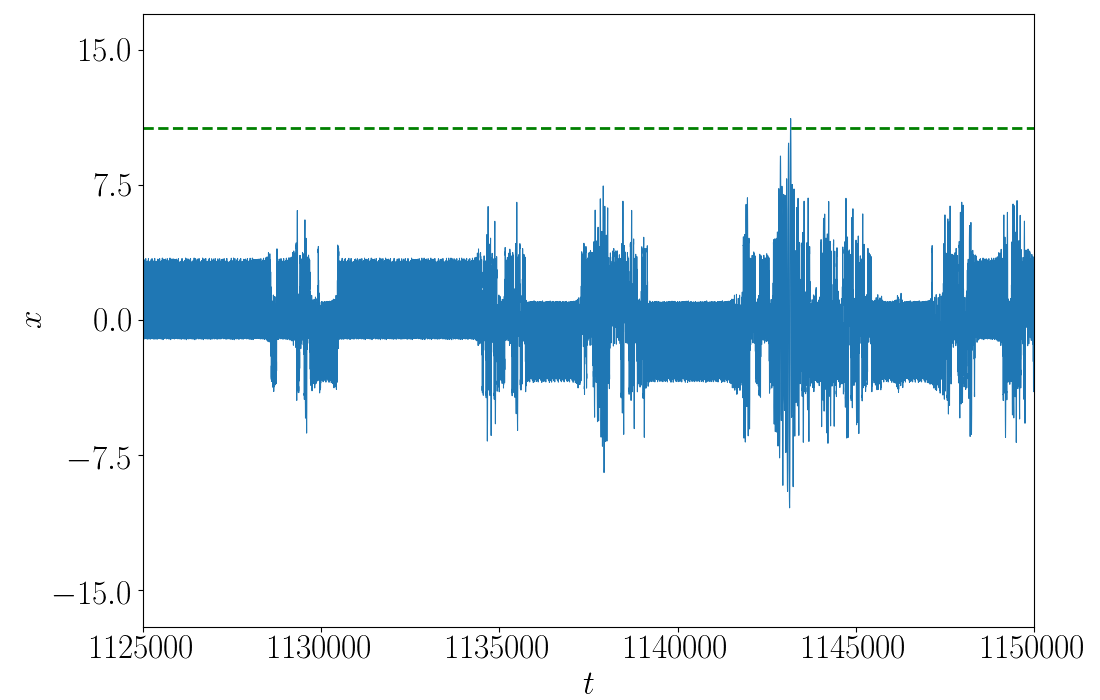}
		\caption{}
		\label{ts1}
	\end{subfigure}
	\begin{subfigure}[b]{0.32\textwidth}
		\centering
		\includegraphics[width=0.7\textwidth]{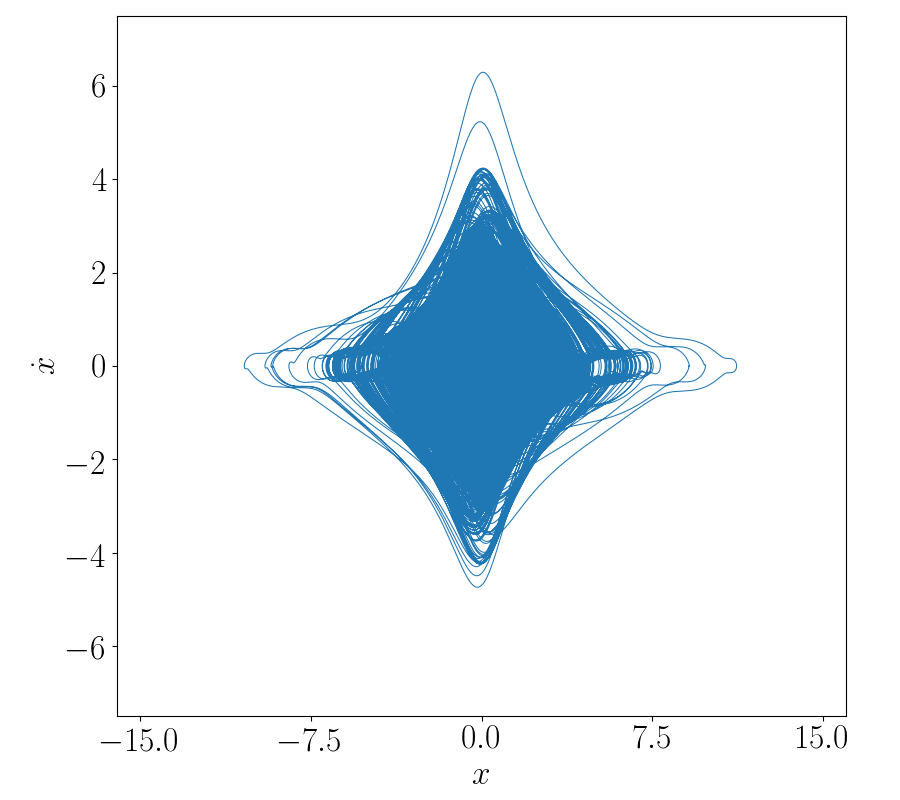}
		\caption{}
		\label{phase1}
	\end{subfigure}
	\begin{subfigure}[b]{0.32\textwidth}
		\centering
		\includegraphics[width=0.9\textwidth]{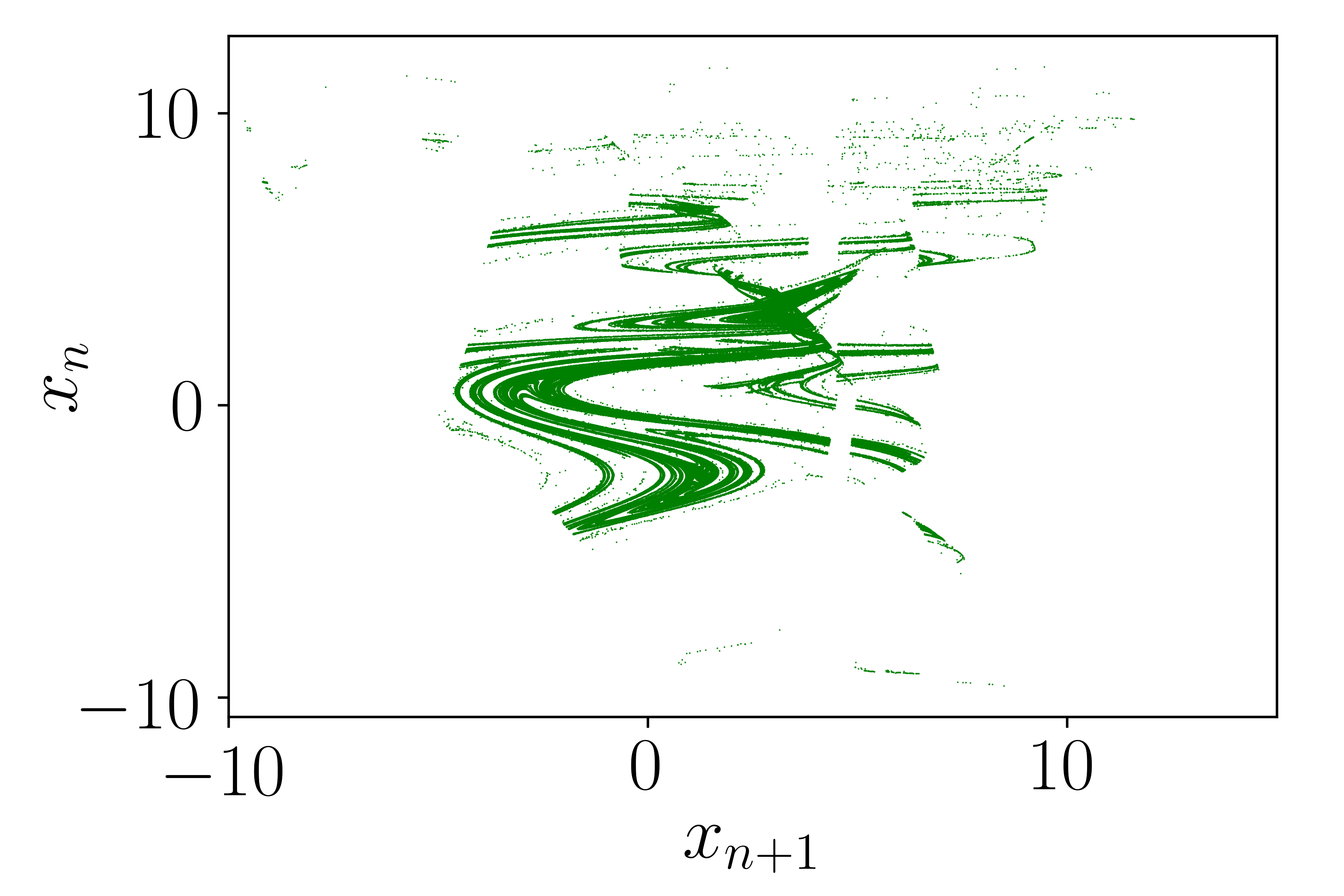}
		\caption{}
		\label{return1}
	\end{subfigure}
	\caption{(a) Time series of system (\ref{delay}) at $f=3.055$. Threshold $x_{ee}$ is shown by green dotted horizontal line. (b) Phase portrait of system (\ref{delay}) and (c) Return map of system (\ref{delay}) at $\epsilon=0.019$.}
	\label{timephase}
\end{figure}

Now we display the time series at $\epsilon=0.019$ in Fig. \ref{ts1}. The threshold $x_{ee}$ is calculated as mentioned in the previous section and is shown by the green dotted horizontal line in Fig. \ref{ts1}. We can see the crossing of the trajectory beyond the threshold confirming the occurrence of extreme events. The corresponding phase portrait is shown in Fig. \ref{phase1}. The excursion can be clearly seen as a large deviation from the bounded chaotic orbit. The chaotic nature of the peaks at $\epsilon=0.019$ is also confirmed by the return map in Fig. \ref{return1}.

\begin{figure}[!ht]
	\begin{center}
		\includegraphics[width=0.8\linewidth]{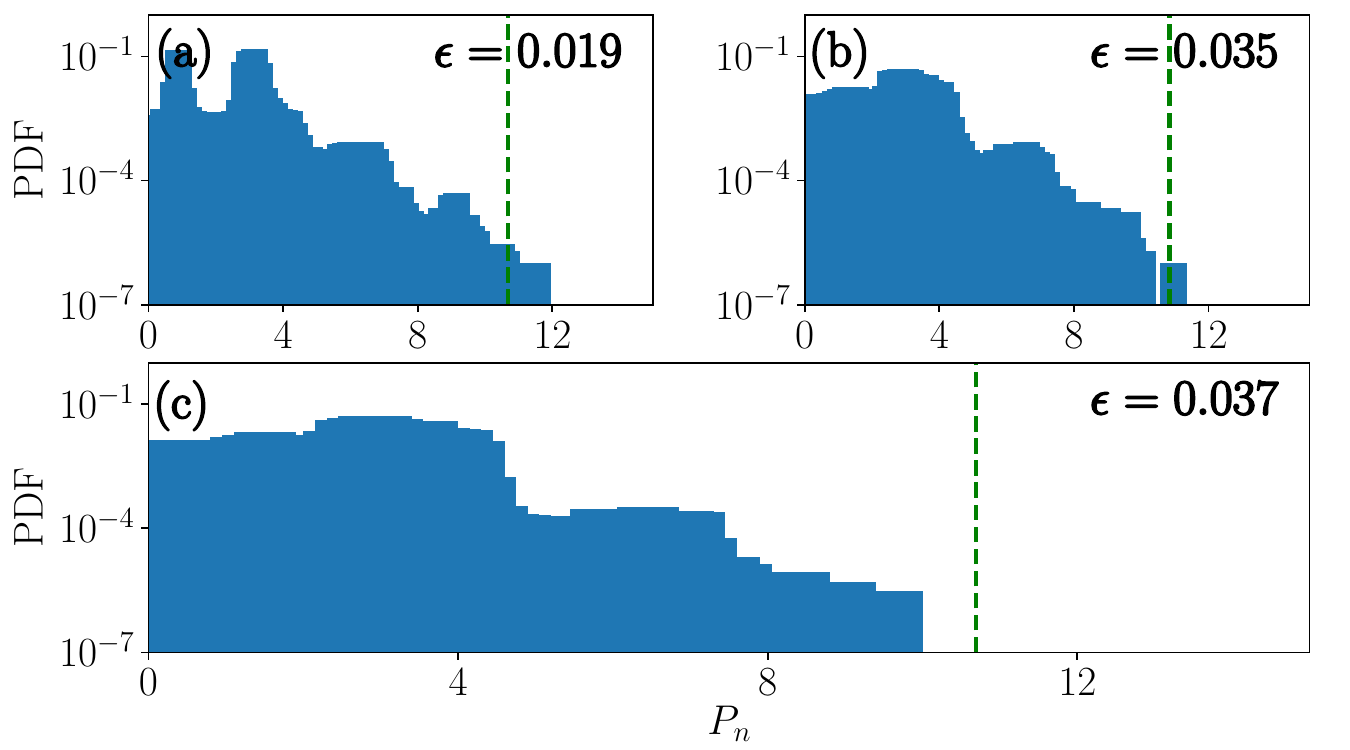}
	\end{center}
	\caption{Plots of the probability distribution function (PDF) of peaks ($P_N$) for system (\ref{delay}) with (a) $\epsilon=0.019$,  (b) $\epsilon=0.035$ and (c)  $\epsilon=0.036$. The qualifier threshold is noted by green dotted vertical line.}
	\label{pdf1}
\end{figure}

Next, we show the emergence and the corresponding suppression of extreme events by probability distribution function (PDF) of peaks in Fig. \ref{pdf1} with (a) $\epsilon=0.019$,  (b) $\epsilon=0.035$ and (c)  $\epsilon=0.036$. We can observe a decrease in both the total number of peaks beyond $x_{ee}$ and its corresponding probability (compare Fig.~\ref{pdf1}(b) with \ref{pdf1}(a)). Finally, when $\epsilon=0.037$, where no extreme events occur, there are no peaks beyond threshold as shown in Fig. \ref{pdf1}(c). Thus when $f=3.055$, when time-delayed feedback is introduced, extreme events get induced in the system (\ref{delay}) and subsequently when the value of $\epsilon$ is increased further, extreme events get suppressed.

\begin{figure}[!ht]
	\begin{center}
		\includegraphics[width=0.7\linewidth]{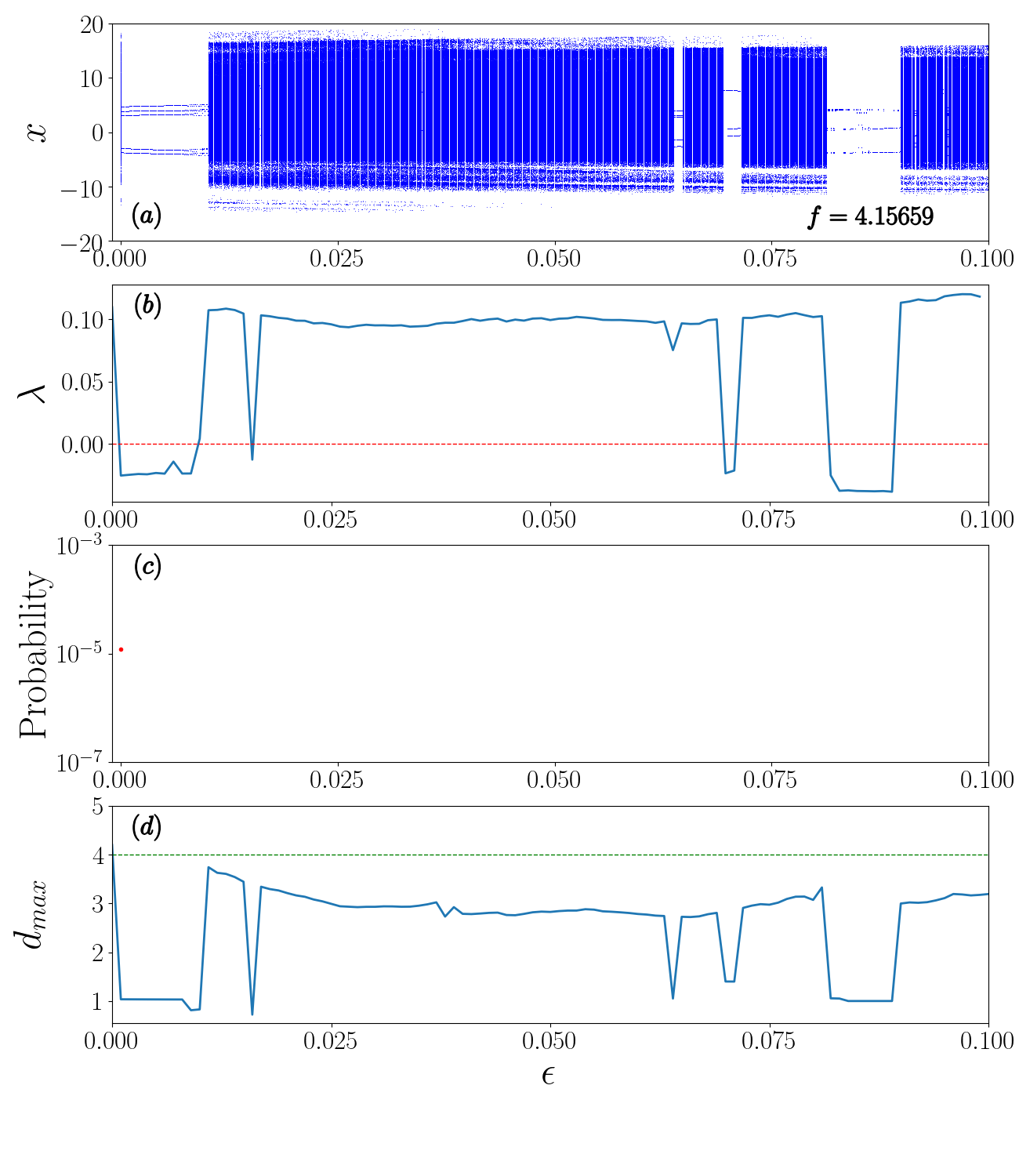}
	\end{center}
	\caption{(a) Bifurcation plot, (b) Largest Lyapunov exponent, (c) Probability plot and (d) $d_{max}$ plot of Eq.~\ref{delay} for $f=4.15659$ for varying $\epsilon$}
	\label{bif4.15659}
\end{figure}

Next for $f=4.15659$, we analyze the effect of $\epsilon$ on the extreme events. The outcome is shown in Fig. \ref{bif4.15659}. It can be observed that immediately on the introduction of time-delayed feedback, chaotic nature of the system is destroyed due to tangent bifurcation. But again on increasing the value of $\epsilon$, period doubling bifurcation occurs bringing back the chaotic nature of the system by a sudden expansion. Similar to the other values of $f$, in the present case also, tangent bifurcation occurs intermittently. The range of $\epsilon$ for which the tangent bifurcation occurs is slightly larger in this case. For several regions of $\epsilon$, chaotic attractor prevails. For larger values of $\epsilon$, reverse period doubling occurs. Even for a very small value of $\epsilon$, extreme events get suppressed completely and no further emergence is seen even for larger values of $\epsilon$. This further confirms that extreme events completely get suppressed. Similar to the previous values of $f$, we plot and confirm the nature of the attractor by evaluating the Lyapunov exponent (see Fig.~\ref{bif4.15659}(b)). The dynamics is largely chaotic with positive Lyapunov exponent in most of the regions. It takes negative values only for the periodic windows. The immediate suppression of extreme events can be visualized from the probability plot shown in Fig.~\ref{bif4.15659}(c), where zero probability prevails after the initial point. This can further be confirmed by the $d_{max}$ plot produced in Fig.~\ref{bif4.15659}(d). Since time-delayed feedback suppressed the extreme events completely, the time series, phase portraits and PDF plots will look similar. Hence, we do not present those plots for $f=4.15659$.

\begin{figure}[!ht]
	\begin{center}
		\includegraphics[width=0.7\linewidth]{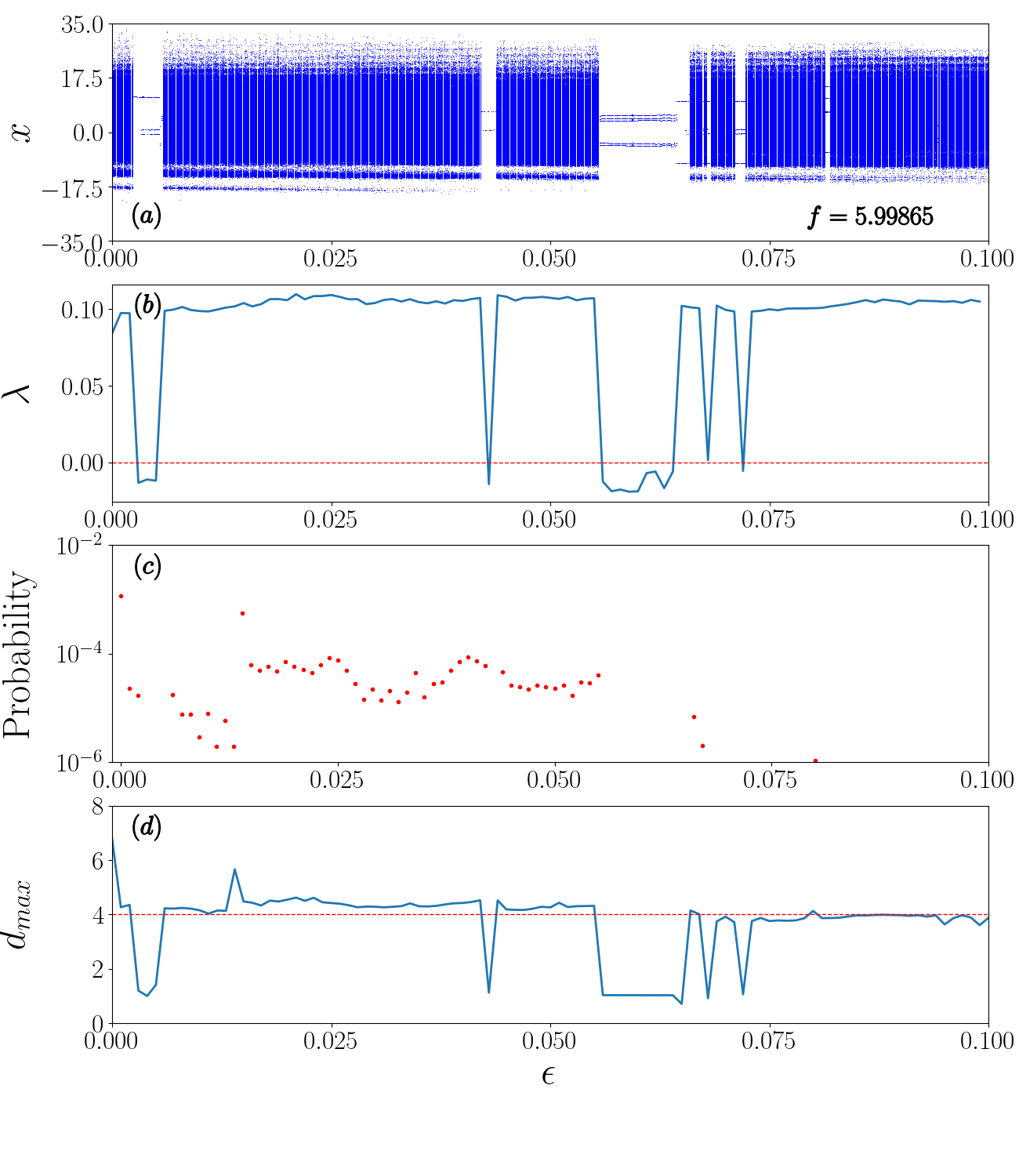}
	\end{center}
	\caption{(a) Bifurcation plot, (b) Largest Lyapunov exponent, (c) Probability plot and (d) $d_{max}$ plot of Eq.~\ref{delay} for $f=5.99865$ for varying $\epsilon$}
	\label{bif5.99865}
\end{figure}

Finally, we investigate the influence of $\epsilon$ when $f=5.99865$. The corresponding results are shown in Fig. \ref{bif5.99865}. From Fig.~\ref{bif5.99865}(a) we can observe that the large sized chaotic attractor prevails throughout the parameter space but destruction and emergence of chaotic nature occur then and there either by tangent bifurcation or through period doubling route. This is again confirmed by calculating the Lyapunov exponent whose outcome is shown in Fig.~\ref{bif5.99865}(b). From Fig. \ref{bif5.99865}(c), we observe that extreme events do not get suppressed immediately as in the case of $f=4.15659$. Rather it completely gets suppressed only for a larger value of $\epsilon=0.080$. This is confirmed by the $d_{max}$ plot shown in Fig.~\ref{bif5.99865}(d) where after complete suppression, the value of $d_{max}$ lies below the value $4$.

\begin{figure}
	\centering
	\begin{subfigure}[b]{0.32\textwidth}
		\centering
		\includegraphics[width=\textwidth]{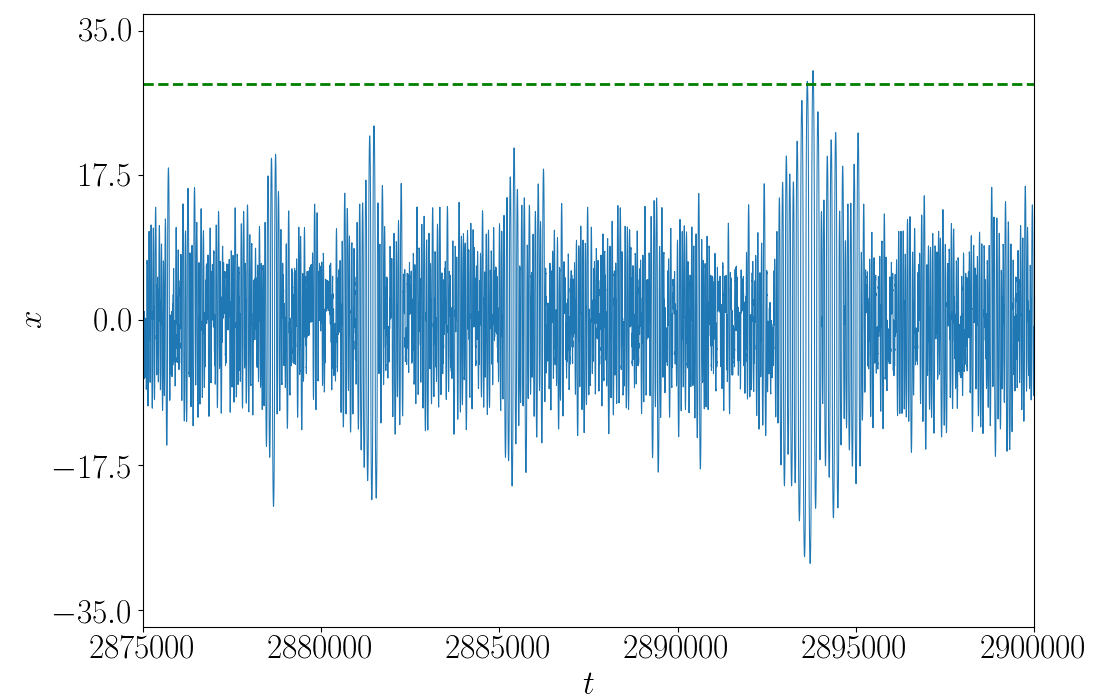}
		\caption{}
		\label{ts2}
	\end{subfigure}
	\begin{subfigure}[b]{0.32\textwidth}
		\centering
		\includegraphics[width=0.7\textwidth]{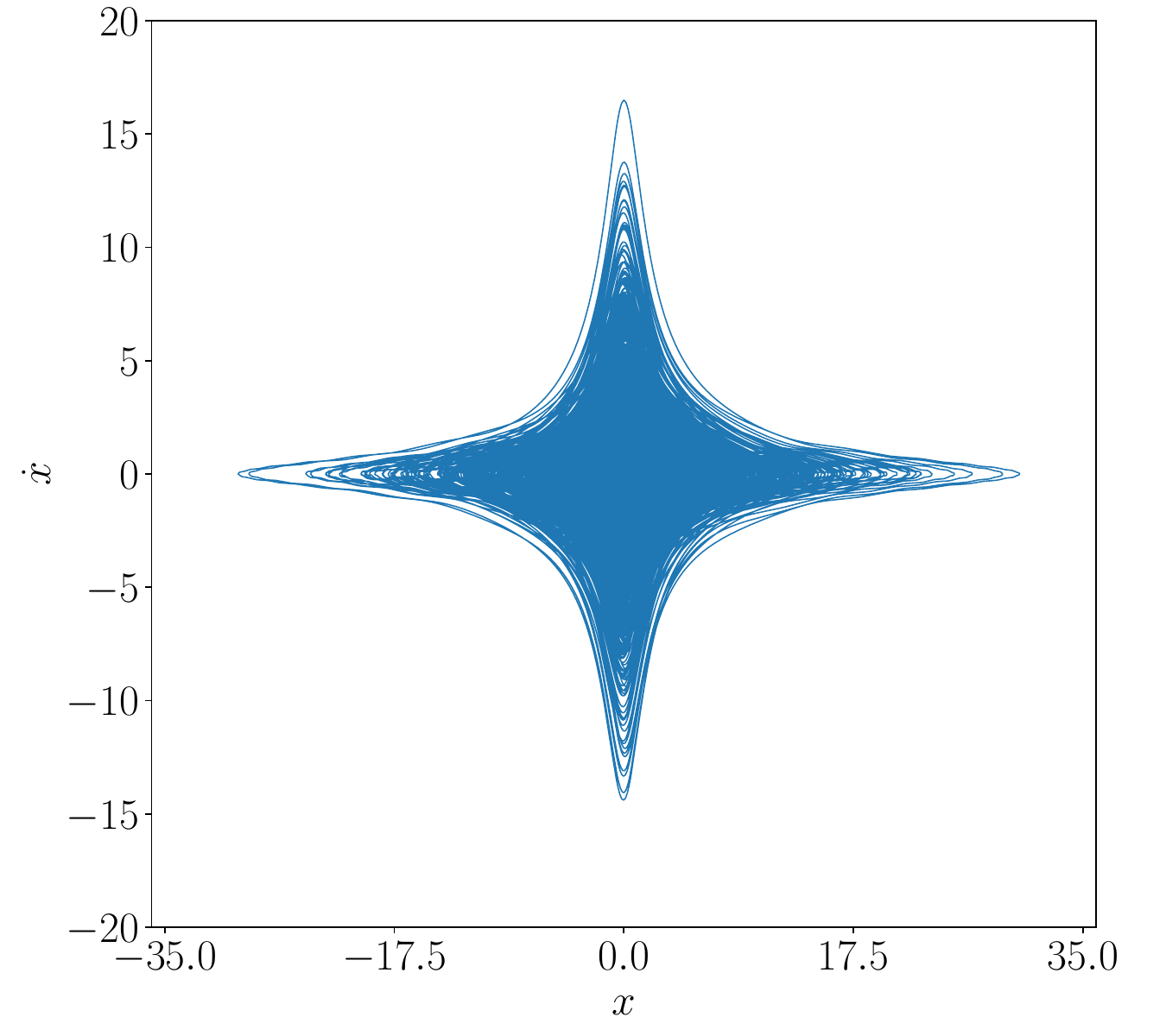}
		\caption{}
		\label{phase2}
	\end{subfigure}
	\begin{subfigure}[b]{0.32\textwidth}
		\centering
		\includegraphics[width=0.9\textwidth]{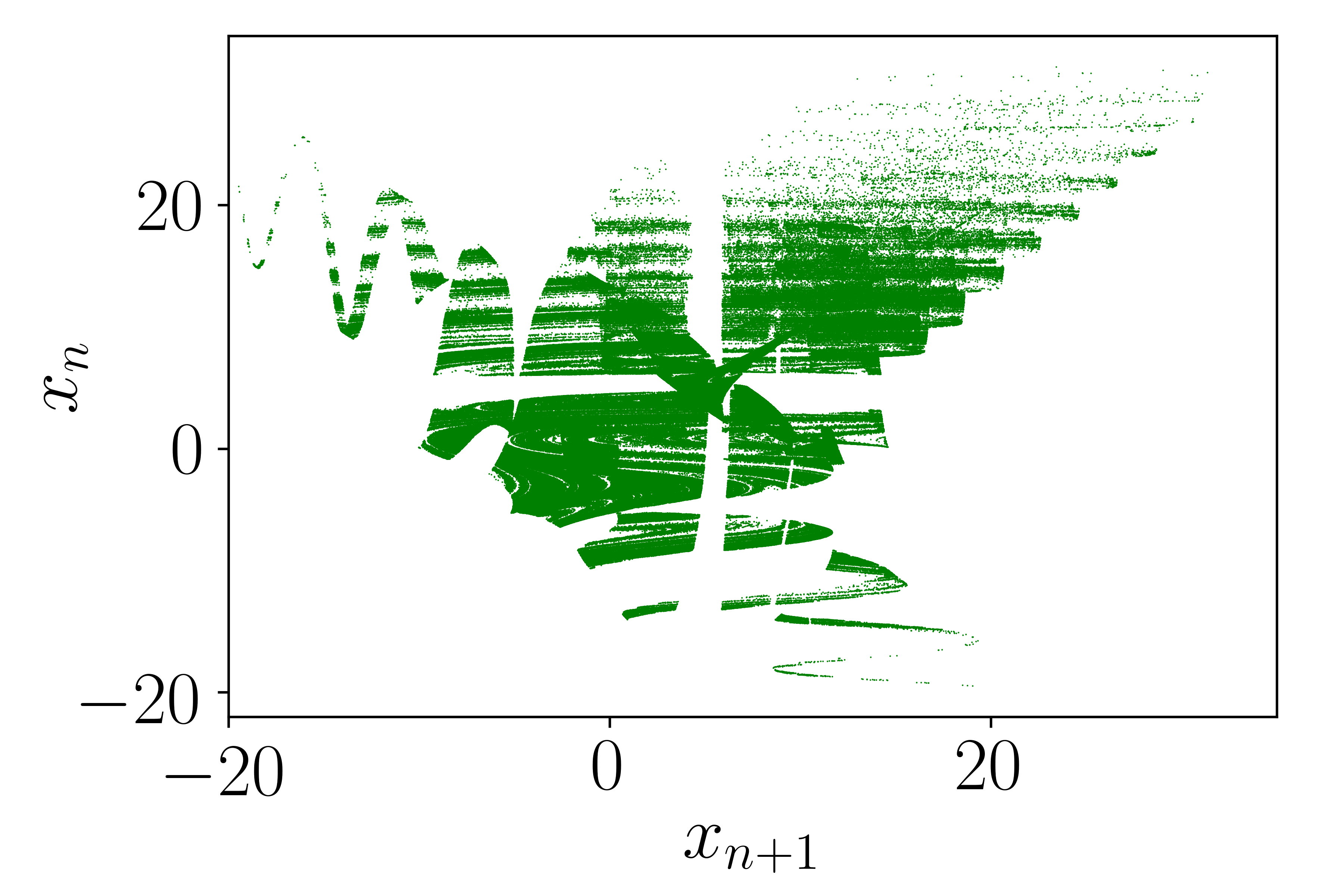}
		\caption{}
		\label{return2}
	\end{subfigure}
	\caption{(a) Time series of system (\ref{delay}) at $f=5.99865$ and $\epsilon=0.24$. Threshold $x_{ee}$ is shown by green dotted horizontal line. (b) Phase portrait and (c) Return map of system (\ref{delay}) at $f=5.99865$ and $\epsilon=0.24$.}
	\label{timephase2}
\end{figure}

The time series of the system at $\epsilon=0.024$ is shown in Fig. \ref{timephase2}(a) and the crossing of the trajectory beyond the threshold can clearly be seen in the given time interval. The corresponding phase portrait is given in Fig. \ref{timephase2}(b). Here also the chaotic nature of the peaks are confirmed by the return map given in Fig. \ref{timephase2}(c).  

\begin{figure}[!ht]
	\begin{center}
		\includegraphics[width=0.8\linewidth]{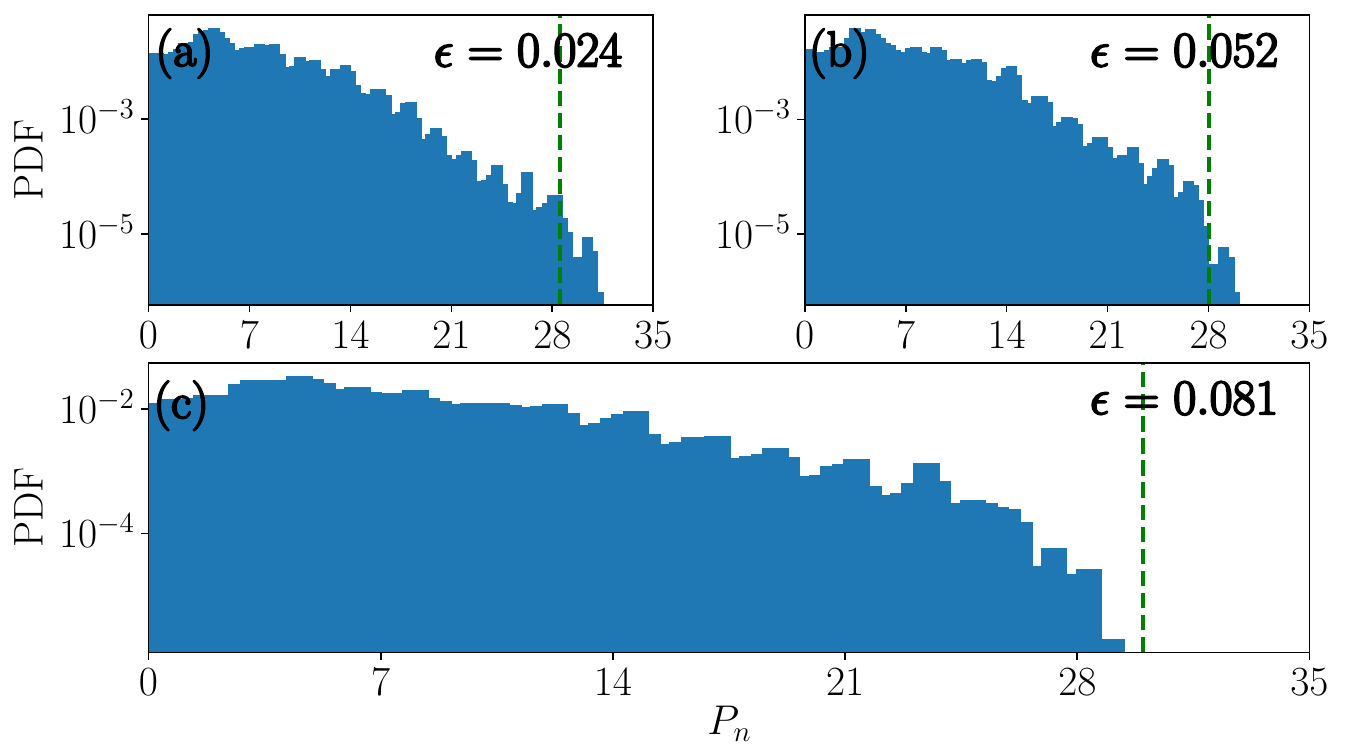}
	\end{center}
	\caption{Plots of the probability distribution function (PDF) of peaks ($P_N$) for system (\ref{delay})with $f=5.99865$ and with (a) $\epsilon=0.024$,  (b) $\epsilon=0.052$ and (c)  $\epsilon=0.081$. The qualifier threshold is noted by green dotted vertical line.}
	\label{pdf2}
\end{figure}

The plots of the PDF of peaks ($P_N$) for the system (\ref{delay}) with (a) $\epsilon=0.024$,  (b) $\epsilon=0.052$, and (c)  $\epsilon=0.081$, are given in Fig.~\ref{pdf2}. The qualifier threshold is noted by green dotted vertical line. Similar to the previous values of $f$, here also we can see that peaks with non-zero probability are present beyond the qualifier threshold for $\epsilon=0.024$ and when $\epsilon=0.052$, the number of peaks with non-zero probabily beyond the threshold decreases. This is shown in Figs. \ref{pdf2}(a) and Fig. \ref{pdf2}(b). Finally, at $\epsilon=0.081$, when there is no occurrence of extreme events, there is no peak beyond the qualifier threshold.

For other values of $f$ for which the system exhibits {\it 1-period, 2-period and 5-period} dynamics, the introduction of time-delay feedback does not produce extreme events.

\section{Influence of positive time-delayed feedback}
\label{sec:5}

\begin{figure}[!ht]
	\begin{center}
		\includegraphics[width=0.7\linewidth]{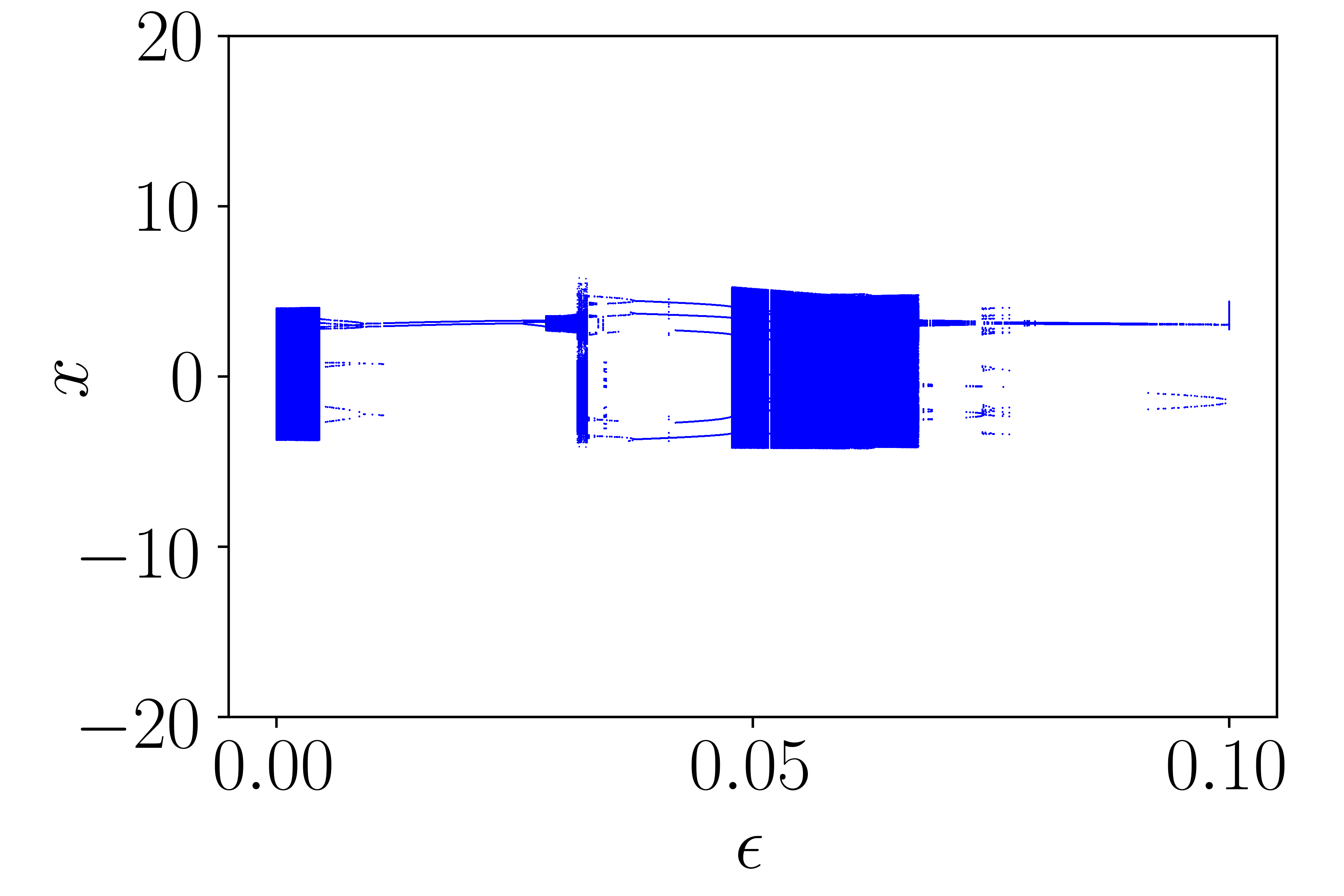}
	\end{center}
	\caption{Bifurcation plot of Eq.~\ref{delay} forr $f=2.7$ for varying $\epsilon$}
	\label{bif2.7a}
\end{figure}

In this section, we investigate the influence of positive time-delayed feedback on the extreme events. Since reproducing and discussing the time series, phase portrait, return map and the PDF plots for this case would be repetitive and exhaustive, we restrict ourselves with the bifurcation, Lyapunov exponent, probability and the $d_{max}$ plots. In the positive time-delayed feedback case, we observe significant changes, both qualitative and in the observation of extreme events. For $f=2.7$, upon comparing the Figs. \ref{bif2.7a} and \ref{bif2.7}, one may note that, in the case of negative time-delayed feedback (see Fig. \ref{bif2.7}),  the chaotic nature of the system prevails, whereas in the present case (see Fig.~\ref{bif2.7a}), the chaotic nature of the system is destroyed in most of the regions. But still we do not observe any occurrence of extreme events in the negative time-delayed feedback. Since no extreme events occurred before and after the influence of time-delayed feedback, we do not present any supporting plots here.

\begin{figure}[!ht]
	\begin{center}
		\includegraphics[width=0.7\linewidth]{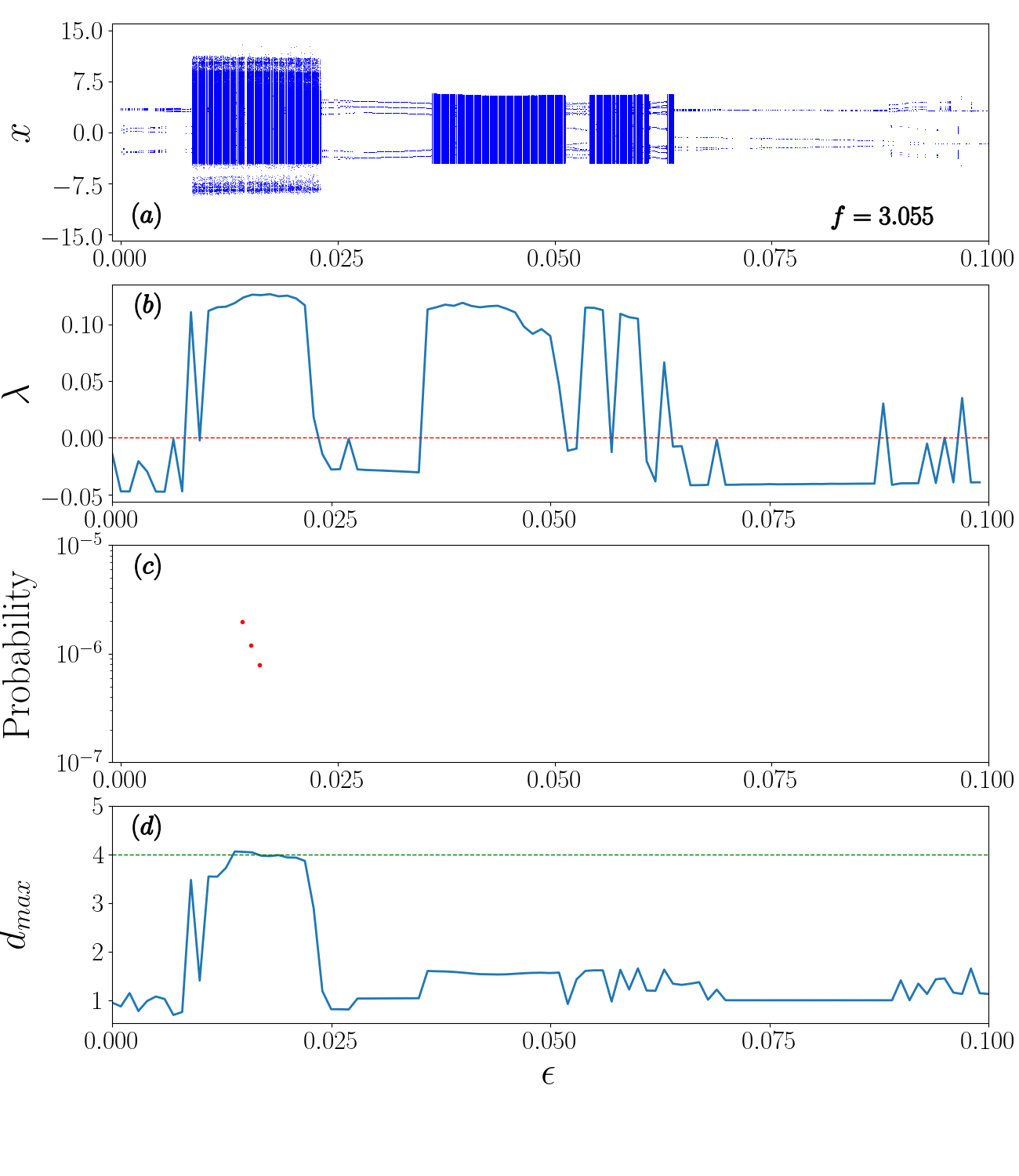}
	\end{center}
	\caption{(a) Bifurcation plot, (b) Largest Lyapunov exponent, (c) Probability plot and (d) $d_{max}$ plot of Eq.~\ref{delay} for $f=3.055$ for varying $\epsilon$}
	\label{bif3.055a}
\end{figure}

For $f=3.055$, we plot the bifurcation, Lyapunov exponent, probability and $d_{max}$ plot respectively in Figs. \ref{bif3.055a}(a), (b), (c) and (d). Similar to the previous case, here also, we observe that the regions of chaotic attractor decreases when compared to the negative feedback case. The suppression of chaotic region is confirmed by the change of Lyapunov exponent from positive to negative, see Fig.~\ref{bif3.055a}(b). As we have seen before, at $f=3.055$ there is no emergence of extreme events in the absence of time-delayed feedback. Now when the time-delayed feedback is introduced and varied, we observe extreme events emerge for a very small range of $\epsilon$, similar to the negative feedback case but the range is much smaller in the case of positive feedback. In the positive feedback case, the extreme events do not occur at the point of tangent bifurcation. Rather they occur in the middle of the expanded chaotic attractor as shown in Fig \ref{bif3.055a}(c). To be precise, we find the emergence occurs only at three different values of $\epsilon$. Only at these points of $\epsilon$ the value of $d_{max}$ crosses $n=4$ and lies well below four for all other values of $\epsilon$ as shown in Fig.~\ref{bif3.055a}(d).

\par
\begin{figure}[!ht]
	\begin{center}
		\includegraphics[width=0.7\linewidth]{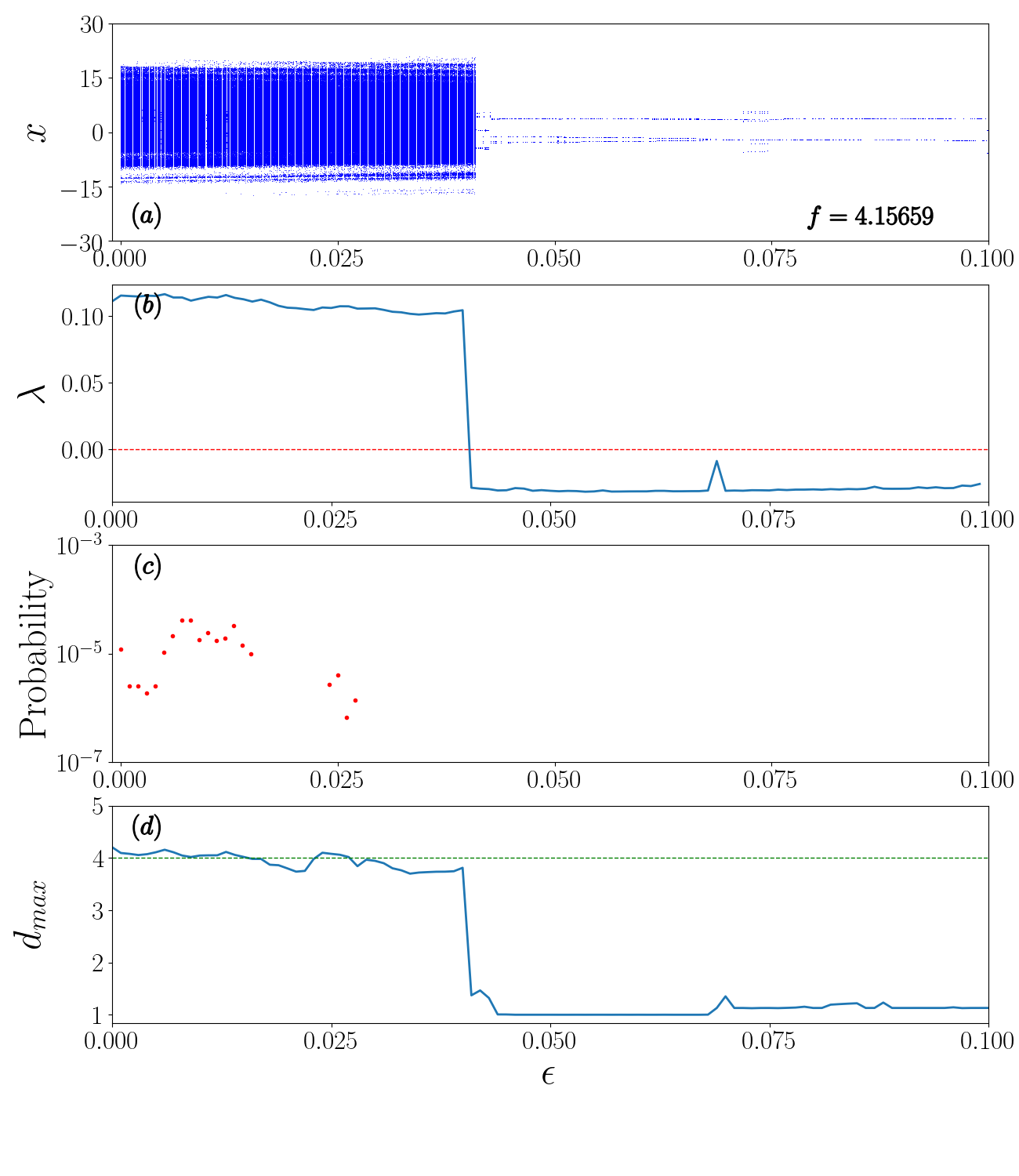}
	\end{center}
	\caption{(a) Bifurcation plot, (b) Largest Lyapunov exponent, (c) Probability plot and (d) $d_{max}$ plot of Eq.~\ref{delay} for $f=4.15659$ for varying $\epsilon$}
	\label{bif4.15659a}
\end{figure}

\begin{figure}[!]
	\begin{center}
		\includegraphics[width=0.7\linewidth]{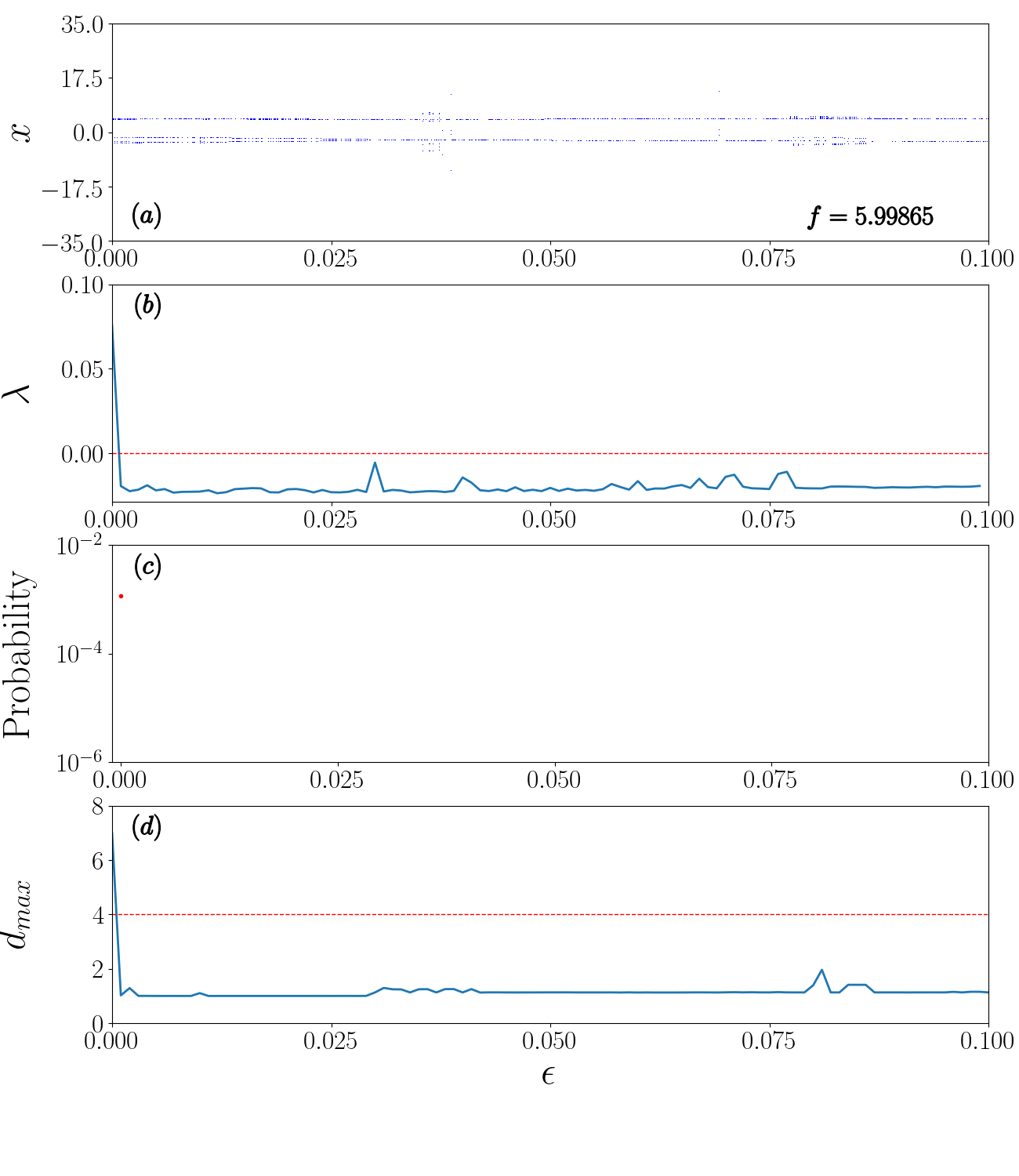}
	\end{center}
	\caption{(a) Bifurcation plot, (b) Largest Lyapunov exponent, (c) Probability plot and (d) $d_{max}$ plot of Eq.~\ref{delay} for $f=5.99865$ for varying $\epsilon$}
	\label{bif5.99865a}
\end{figure}

 Next we analyze the effect of positive $\epsilon$, for $f=4.15659$ and show the results in Fig. \ref{bif4.15659a}. This is the point where a significant changes occur not only in the dynamics of the system but also in the occurrence of extreme events.  The suppression of chaos is confirmed by the transition of the Lyapunov exponent from positive to negative at $\epsilon=0.041$ as shown in Fig.~\ref{bif5.99865a}(b). Figure \ref{bif4.15659}, reveals that introduction of negative time-delayed feedback at $f=4.15659$ has suppressed the occurrence of extreme events immediately. Surprisingly, we found that such a sudden suppression does not occur under the influence of positive time-delayed feedback. Rather, complete suppression occurs only for a larger value of $\epsilon = 0.041$. It can be seen that, in the present case, the suppression of extreme events and the suppression of chaos occurs only at two different values of $\epsilon$.

Finally, for $f=5.99865$, we present the changes in the dynamics of the system through the bifurcation diagram in Fig.~\ref{bif5.99865a}(a) and the changes in the occurrence of extreme events in Fig.~\ref{bif5.99865a}(b). Further, we also produce the corresponding Lyapunov exponent and $d_{max}$ respectively in Figs. \ref{bif5.99865a}(c) and (d). In the positive feedback case also, we observe a drastic change in the dynamics similar to the case of $f=4.15659$. When the time-delayed feedback was negative, we observe a complete suppression of extreme events only at a larger value of $\epsilon$. The suppression occurred gradually. Now in the positive time-delayed feedback case, we observe an opposite dynamics. The chaotic attractor completely gets destroyed and the extreme events completely gets suppressed immediately after the introduction of a very small time-delayed feedback strength. The suppression of chaos can be confirmed by the transition of Lyapunov exponent to negative values. The suppression of extreme events are first substantiated using the probability plot and then using the $d_{max}$ plot.

\section{Effect of time-delay feedback on overall dynamics of the system}
\label{sec:6a}

\begin{figure}[!ht]
	\begin{center}
		\includegraphics[width=0.85\textwidth]{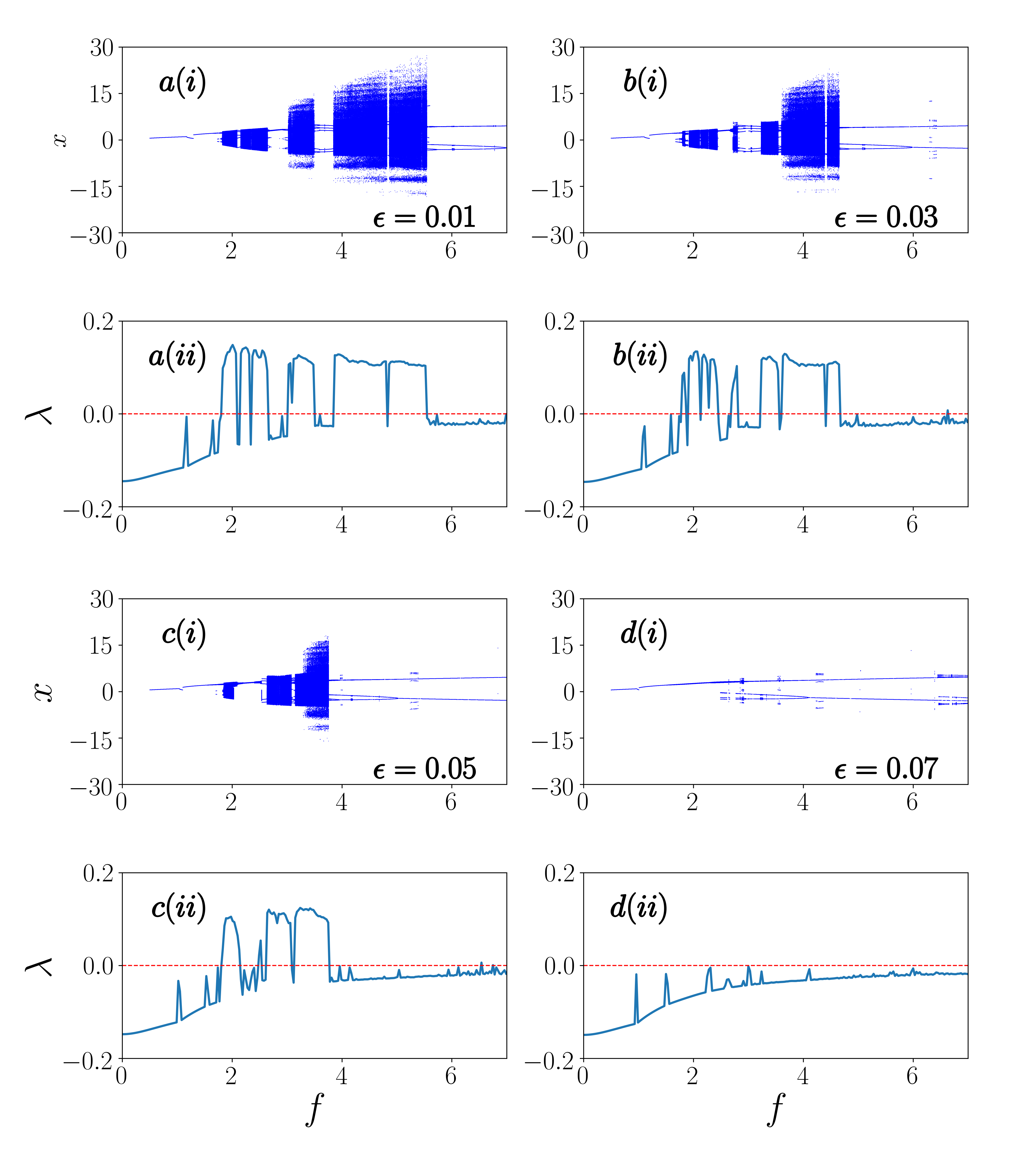}
	\end{center}
	\caption{Subplots with indices (i) represent the bifurcation diagram and plots with indices (ii) correspond to the Lyapunov exponent for the system (\ref{delay}) for various values of $\epsilon$, namely $\epsilon=0.01$ for subplots $a$, $\epsilon=0.03$ for subplots $b$, $\epsilon=0.05$ for subplots $c$ and $\epsilon=0.07$ for subplots $d$.}
	\label{full_peps}
\end{figure}

Now we consider the time-delayed feedback to be positive and vary the corresponding feedback strength and study the changes that occur in the dynamics of the system. On increasing the feedback strength systematically from $\epsilon=0.0$ in the order 0.01, we find that the chaotic attractor in the system begins to destroy by the movement of the periodic orbits (present beyond $f=6.0$ at $\epsilon=0.0$) towards left. We find that for every increase in the value of $\epsilon$, this periodic orbit further moves towards the left and at a particular value of $\epsilon=0.07$, the chaotic attractor completely gets suppressed. This systematic destruction of chaos can be seen from Fig.~\ref{full_peps}. Sub-plots in Fig.~\ref{full_peps} with indices (i) represent the bifurcation diagram and plots with indices (ii) correspond to Lyapunov exponent of the system (\ref{delay}) for various values of $\epsilon$, namely $\epsilon=0.01$ for subplots $a$, $\epsilon=0.03$ for subplots $b$, $\epsilon=0.05$ for subplots $c$ and $\epsilon=0.07$ for subplots $d$. For $\epsilon=0.01$, the value of the transition of chaotic attractor to periodic orbit has moved leftwards to $f=5.52$. When $\epsilon=0.05$, the point has moved further leftwards ($f=4.65$). On futher increasing $\epsilon$ to $0.07$, the chaotic attractor completely gets destroyed with only periodic orbit getting prevailed. All these were confirmed through the Lyapunov exponent. For chaotic attractor, the value of the largest Lyapunov exponent is positive and it is negative for periodic orbits. We note here that as we increase the value of $\epsilon$, the point of transition of the Lyapunov exponent from positive to negative in the right end of the parameter space ($f$) moves leftwards. Finally, when $\epsilon=0.07$, there is not even a single positive Lyapunov exponent. This confirms that the chaotic attactor has been destroyed completely due to the positive time-delayed feedback.

\begin{figure}[!ht]
	\begin{center}
		\includegraphics[width=1.0\textwidth]{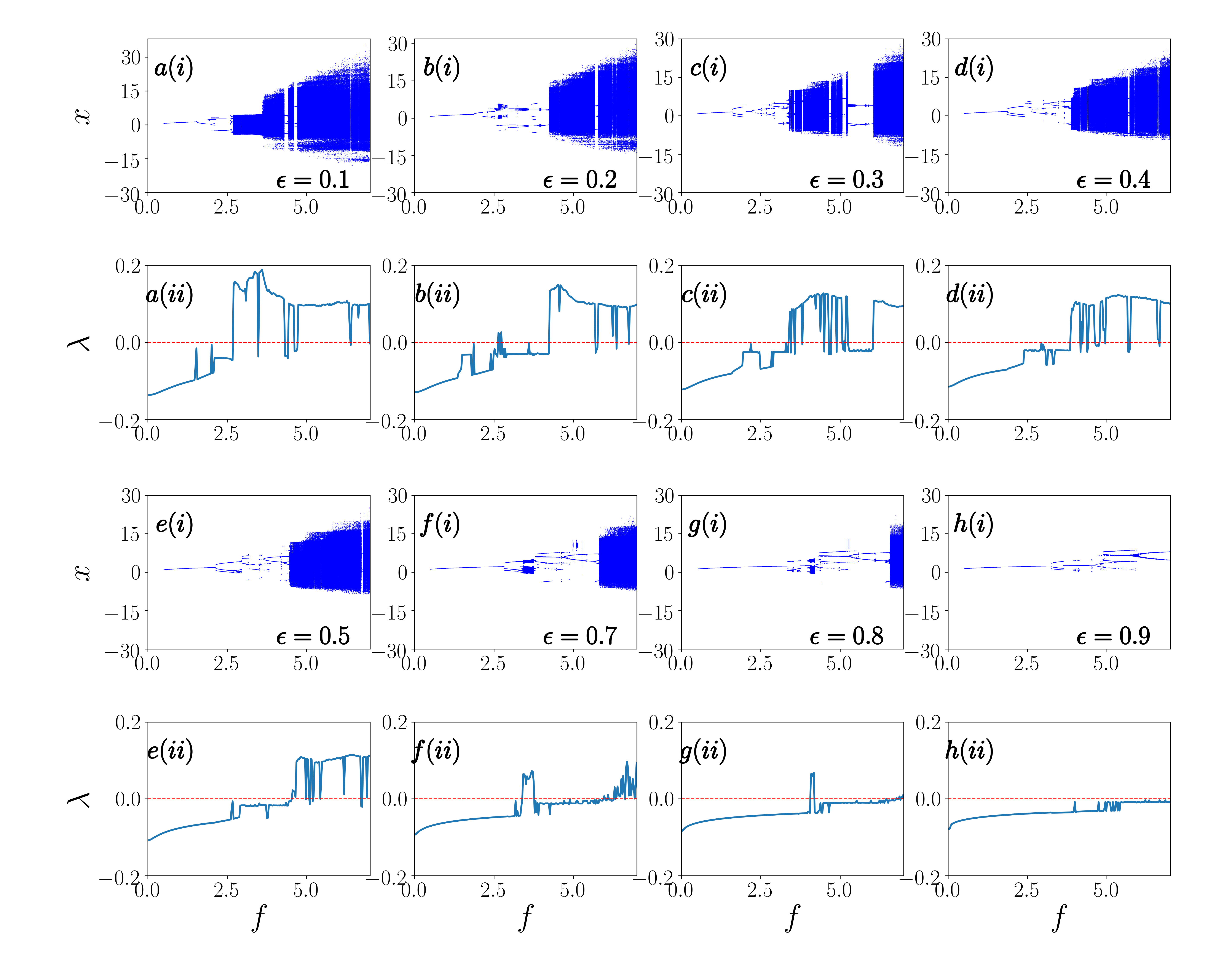}
	\end{center}
	\caption{Subplots with indices (i) represent the bifurcation diagram and plots with indices (ii) correspond to the Lyapunov exponent for the system (\ref{delay}) for various values of $\epsilon$, namely $\epsilon=0.1$ for subplots $a$, $\epsilon=0.2$ for subplots $b$, $\epsilon=0.3$ for subplots $c$ and $\epsilon=0.4$ for subplots $d$, $\epsilon=0.5$ for subplots $e$, $\epsilon=0.7$ for subplots $f$, $\epsilon=0.8$ for subplots $g$ and $\epsilon=0.9$ for subplots $h$.}
	\label{full_neps}
\end{figure}

Now we check the effect of negative time-delayed feedback strength on system (\ref{delay}). Here also, we find that while increasing the value of $\epsilon$, the chaotic attractor suppresses. The nature of suppression is shown in Fig.~\ref{full_neps}. It is evident from the figure that the way in which suppression occurs in the present case is different from that of the positive time-delayed feedback case. Initially, for $\epsilon=0.1$, the chaotic nature of the system increases throughout the parameter space beyond $f=5.0$. When $\epsilon$ is increased to $0.2$, the chaotic attractor present between $f=2.7$ and $f=4.26$ gets destroyed into periodic orbits. Upon further increasing the value of the parameter $\epsilon$ to $0.3$, the dynamics changes completely as shown in Fig.~\ref{full_neps}c(i). Now when the value of $\epsilon$ is changed to $0.4$, any chaotic attractor that present before $f=3.87$ gets completely converted into periodic orbit. Chaotic attractor prevails only beyond $f=3.87$ which are interspersed with periodic windows. From this value of $\epsilon$ onwards, on further inreasing the value of $\epsilon$, we observe that the point of origination of chaotic attractor (which was previously at $f=3.87$ for $\epsilon=0.4$) moves towards right side in the parameter space. In other words the range of the values for which the chaos occurs decreases on increasing the value of $\epsilon$ further. Finally, at $\epsilon=0.9$, the chaos suppresses completely and the sytem exhibits only periodic orbits in the entire parameter space. All these results are confirmed using Lyapunov exponents.

\section{Discussion}
\label{sec:6}
\begin{figure}
	\centering
	\begin{subfigure}[b]{0.35\textwidth}
		\centering
		\includegraphics[width=\textwidth]{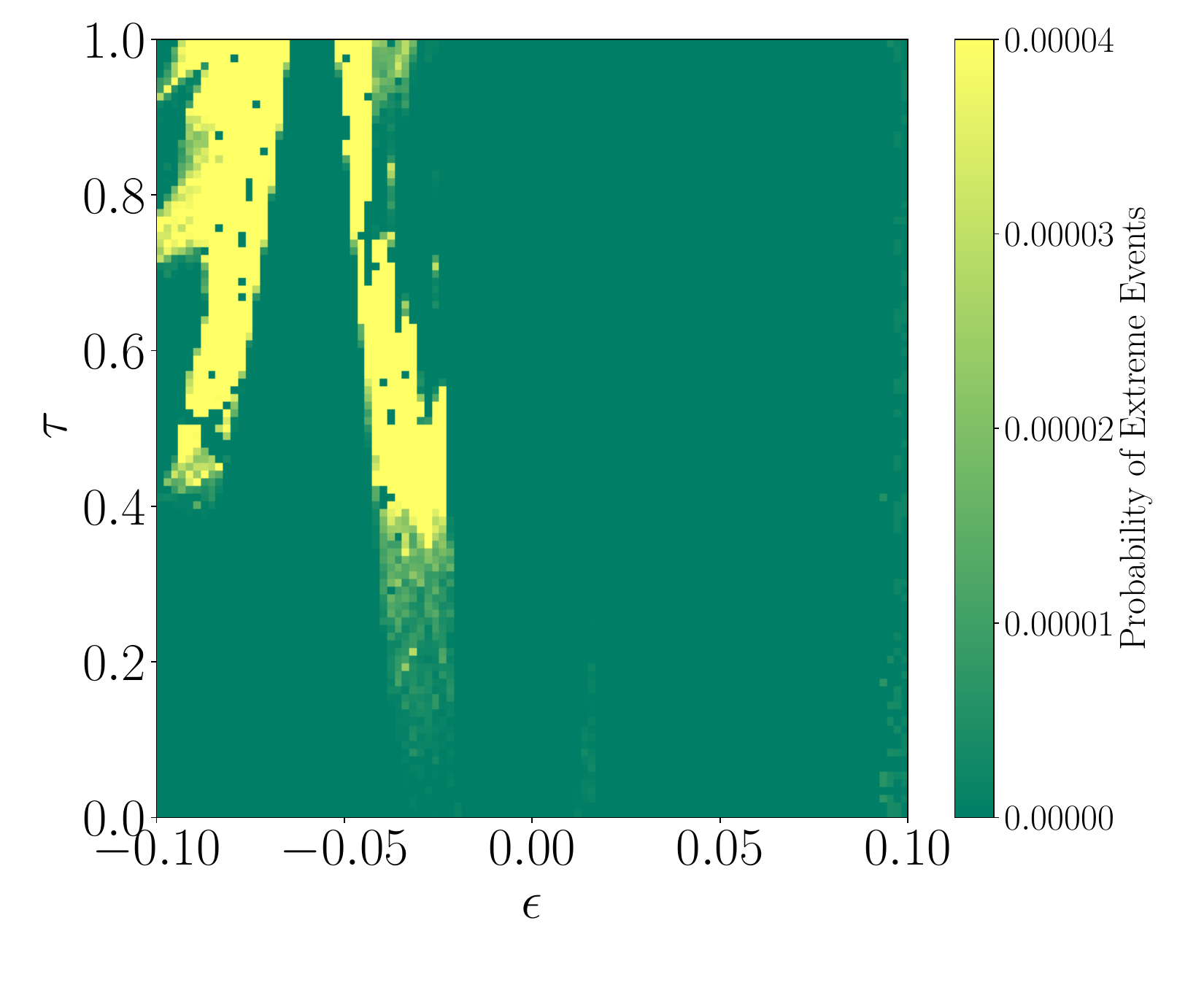}
		\caption{$f=3.055$}
		\label{2phase1}
	\end{subfigure}
	\begin{subfigure}[b]{0.35\textwidth}
		\centering
		\includegraphics[width=\textwidth]{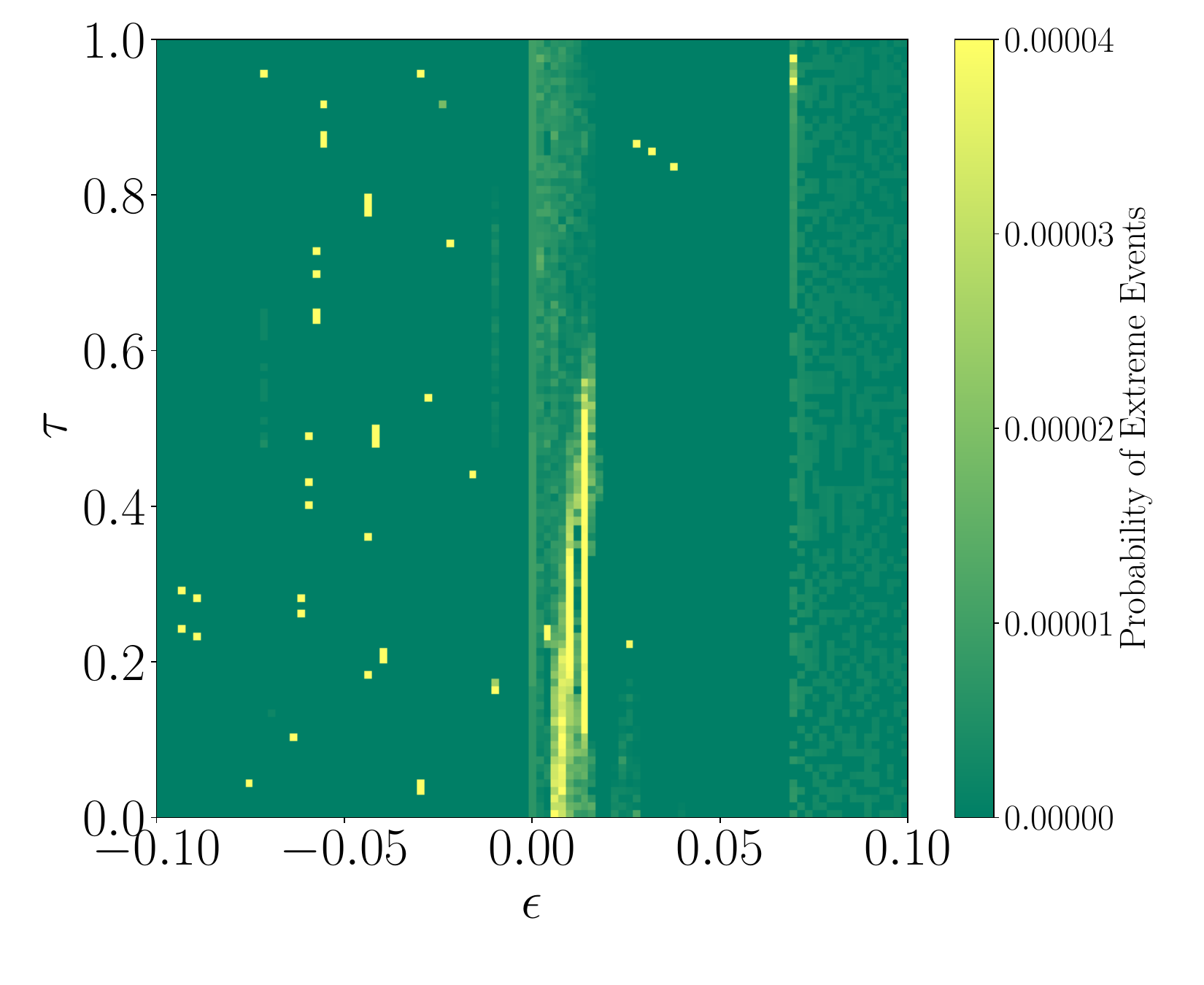}
		\caption{$f=4.15659$}
		\label{2phase2}
	\end{subfigure}
	\hfill
	\begin{subfigure}[b]{0.35\textwidth}
		\centering
		\includegraphics[width=\textwidth]{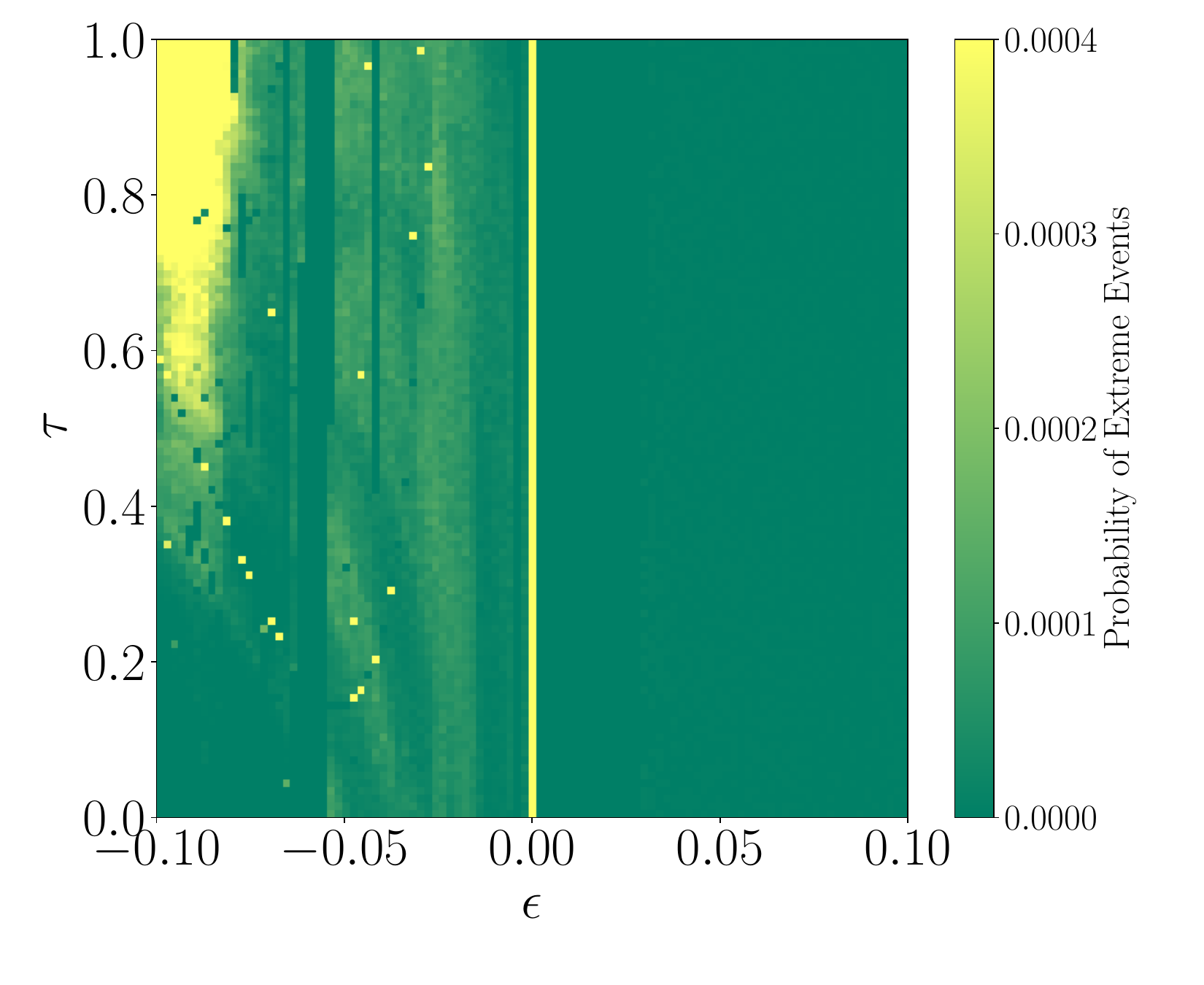}
		\caption{$f=5.99865$}
		\label{2phase3}
	\end{subfigure}
	\caption{Two parameter probability diagram in the $(\epsilon-\tau)$ plane for three different values of $f$.}
	\label{2phase}
\end{figure}

In this section, we consolidate the results of the effect of time-delayed feedback on extreme events and then discuss the results pertaining to the suppresion of extreme events in a nutshell. In Fig.~\ref{2phase}, we present the two parameter probability diagram in the $(\epsilon-\tau)$ plane for three values of $f$, namely $f=3.055$, $f=4.15659$ and $f=5.99865$.

From Fig.~\ref{2phase}, we can infer that extreme events are suppressed in an excellent manner when the time-delayed feedback is positive whereas for the negative feedback case, the suppression is confined to a few parameter values and that too only for low values of $\tau$. In Sec.~3, we found that suppression of chaos happens only for a very small value of the time-delayed feedback strength with $\tau=0.1$. In the two parameter diagram also, we can observe a similar suppression of extreme events for $\tau>0.1$. 

The obtained results of extreme events due to the influence of time-delayed feedback are consolidated in the Table.~{\ref{table1}}. From the outcome, we can observe that in both the negative and positive feedback cases, extreme events mainly occur at the point of expansion of chaotic attractor and continue to get suppressed either immediately or lately in the parameter space. In both the cases, the introduction of time-delay induces a collision of periodic orbit with the chaotic attractor due to which periodic windows occur often. In particular, when time-delayed feedback is introduced in the system, we find that the system exhibits three kinds of properties, that is $(i)$ no extreme events occur both before and after the introduction of time-delay feedback, $(ii)$ extreme events are induced by the time-delay feedback and $(iii)$ extreme events suppressed by the time-delay feedback. These three natures are displayed both under negative and positive feedback. Also from the bifurcation plots that we have presented in the previous two sections, it can be seen that destruction of the chaotic attractor occurs more for the positive feedback case. Overall, the places where extreme events occur is fewer in number in the postive feedback case than that of the other case.

\begin{table}[!ht]
	\begin{center}
		\begin{tabular}{|c|c|c|c|}
			\hline
			\textbf{S. No.}  & \textbf{Value of $f$} & \textbf{Effect of negative} & \textbf{Effect of positive} \\ & & \textbf{time-delayed feedback} & \textbf{time-delayed feedback} \\ \hline\hline\hline
			1 & 2.7  & Neither emergence nor & Neither emergence \\
			& & suppression of extreme events & suppression of extreme events \\
			\hline
			2 & 3.055  & Emergence of extreme events  & Emergence of extreme events \\
			& & for few values of $\epsilon$ & in a very short window of $\epsilon$ \\
			\hline
			3 & 4.15659  & Immediate suppression of &  Suppression of extreme events \\
			& & extreme events & for a larger values of $\epsilon$ \\
			\hline
			4 & 5.99865  & Suppression of extreme events &  Immediate suppression of \\
			& & for a larger values of $\epsilon$ & extreme events \\
			\hline
		\end{tabular}
		\caption{Influence of time-delayed feedback on extreme events for various values of $f$ in Eq.~(\ref{delay}).}
		\label{table1}
	\end{center}
\end{table}

From our analysis, we conclude that extreme events appearing in (\ref{delay}) can be completely removed from the system by positive time-delayed feedback and for a few paramter regions in the negative time-delayed feedback.

\section{Conclusion}
\label{sec:7}

In this work, we have studied the influence time-delayed feedback in a non polynomial system with velocity dependent potential. First, we investigated the effect of time-delay on extreme events. For that we have considered four different values of $f$ for which the system (\ref{delay}) exhibits {\it periodic, period doubling and sudden expansion of chaotic attractor}. Out of these four values of $f$, for the first two values the system does not exhibit extreme events without time-delayed feedback. These two values of $f$ were chosen to check wether extreme events emerge under the influence of time-delay. For the last two parameter values the system exhibits extreme events even in the absence of time-delayed feedback. Further, values of $f$ where no emergence of extreme events both before and after the introduction of time-delay feedback are also observed. This fact can be seen both in the negative and positive feedback cases. One of the most significant results of our study is that extreme events gets suppressed in a larger manner in the system (\ref{delay}) when the feedback is positive. We can see that chaos prevails even after the suppression of extreme events. These are validated using Lyapunov exponents, where the probability of the occurrence of extreme events are zero but the Lyapunov exponent is positive. Further, the suppression of extreme events have been confirmed by the disappearance of long tail beyond the threshold value through the peak PDF plots. Further the suppression was also confirmed using Probability plots and $d_{max}$ plots. All confirm the suppression of extreme events as discussed. Finally, we analysed how the overall dynamics of the considered system changes for both positive and the negative values of the time-delayed feedback. We found that when the feedback is positive the chaotic nature of the system is supressed for a very small value of $\epsilon$ itself. When the feedback is negative, chaotic nature of the system is suppressed for a larger values of $\epsilon$. The suppression of chaos is confirmed through Lyapunov exponent. Beyond a certain value of the time-delay feedback strength, chaos and hence extreme events gets completely suppressed from the system. Thus, depending on the requirement under consideration, either extreme events can alone be suppressed without the destruction of chaotic attractor or both extreme events and chaotic nature can be eliminated from the system.
\section*{Acknowledgements}

SS  thanks  the  Department  of  Science and  Technology (DST), Government of India, for support through INSPIRE Fellowship (IF170319).  The work of AV forms a part of a research project sponsored by DST under the Grant No. EMR/2017/002813. The work of PM forms parts of sponsored research projects by Council of Scientific and Industrial Research (CSIR), India (Grant No. 03(1422)/18/EMR-II), and Science and Engineering Research Board (SERB), India (Grant No. CRG/2019/004059). The work of MS forms a part of a research project sponsored by CSIR, India under the Grant No. 03(1397)/17/EMR-II. M.S. also acknowledges MHRD RUSA 2.0 (Physical Sciences) and DST-PURSE (Phase-II) Programmes for providing financial support.


%
\end{document}